\documentclass[12pt]{article}
\usepackage[bottom]{footmisc}
\usepackage[utf8]{inputenc}
\usepackage{authblk}
\usepackage{epsfig}
\usepackage{amssymb}
\usepackage{amsfonts}
\usepackage[mathscr]{euscript}
\usepackage{amsbsy}
\usepackage[all]{xy}
\usepackage{amsmath}
\usepackage{enumerate}
\usepackage{float}

 \usepackage{enumerate}
 \usepackage{hyperref}

\usepackage{amssymb,amscd}
\usepackage{amsmath,amsthm}
\usepackage{xspace}
\usepackage{url}
\newcommand{\del}{\partial}
\newcommand{\ov}{\overline}

\usepackage{epsfig}
\usepackage{amssymb}
\usepackage{amsfonts}
\usepackage{amsbsy}
\usepackage[all]{xy}
\usepackage{amsmath}
\usepackage{enumerate}

\usepackage{amssymb,amscd}
\usepackage{mathrsfs}
\usepackage{amsmath,amsthm}
\usepackage{xspace}
\usepackage{url}

\numberwithin{equation}{section}

\newcommand{\bea}{\begin{eqnarray}\displaystyle}
\newcommand{\eea}{\end{eqnarray}}

\title{Holomorphic Surface Defects in Four-Dimensional Chern-Simons Theory}

\author{Ahsan Z. Khan}

\affil{School of Natural Sciences \\ Institute for Advanced Study \\ Einstein Drive, Princeton NJ 08540 }

\begin{document}

\maketitle 

\begin{abstract} We derive the framing anomaly of four-dimensional holomorphic-topological Chern-Simons theory formulated on the product of a topological surface and the complex plane. We show that the presence of this anomaly allows one to couple four-dimensional Chern-Simons theory to holomorphic field theories with Kac-Moody symmetry, where the Kac-Moody level $k$ is critical $k=-h^{\vee}$. Applying this result to a holomorphic sigma model into a complex coadjoint orbit, we derive that four-dimensional Chern-Simons theory admits holomorphic monodromy defects.\end{abstract}

\tableofcontents

\section{Introduction} Four-dimensional Chern-Simons theory is a gauge theory that lives on the product of two surfaces $\Sigma \times C$\footnote{Typically one takes $\Sigma$ to be a flat surface such as $\mathbb{R}^2$, and $C$ to be either the complex plane $\mathbb{C}$, the cylinder $\mathbb{C}^*$ or an elliptic curve $E$. In this paper we will restrict ourselves mostly to $C = \mathbb{C}$.}. Letting $(x,y)$ denote local coordinates on $\Sigma$ and $(z, \bar{z})$ denote local complex coordinates on $C$, the fundamental field of four-dimensional Chern-Simons theory is a partial connection $$A_x \text{d}x + A_y \text{d}y + A_{\bar{z}} \text{d}\bar{z}$$ for a complex Lie group $G$. The action of the theory is \bea S = \frac{1}{2\pi} \int_{\Sigma \times C} \omega \wedge \text{CS}(A),\eea where $\text{CS}(A)$ denotes the Chern-Simons three-form defined with respect to an invariant pairing on $\mathfrak{g} = \text{Lie}(G)$, and $\omega$ is a closed, holomorphic one-form on $C$. The theory has a mixed topological-holomorphic nature. It is topological along $\Sigma$ and holomorphic along $C$. 
 
\paragraph{} There is a remarkable relationship between this holomorphic-topological version of Chern-Simons theory in four dimensions and integrability in two dimensions. The extended operators (defects) of four-dimensional Chern-Simons theory play an important role in this relationship. A number of them have been studied in the literature. They include

\begin{itemize}
\item Wilson lines supported along a line in the topological surface $\Sigma$ and a fixed point in $C$ \cite{Costello:2013zra, Costello:2017dso, Costello:2018gyb}. When $\Sigma = \mathbb{R}^2$ and $C = \mathbb{C}$ it was shown how the amplitude associated to two crossed Wilson lines leads to the rational R-matrix associated to $\mathfrak{g}$. That the amplitude defined this way gives a solution of the Yang Baxter equation follows from the holomorphic-topological nature of the theory. 
\item `t Hooft defects supported along a line in $\Sigma$ and a point in $C$ \cite{Costello:2021zcl}. These are shown to be related to the theory of $\mathbf{Q}$-operators \cite{Baxter:1972hz} in integrable spin chains. 
\item Surface defects placed along $\Sigma$ and a fixed point in $C$ \cite{Costello:2019tri}. The surface defects include ``order" defects, where the defect theory is some two-dimensional Lagrangian field theory with $G$-symmetry such as a free fermion or $\beta \gamma$ system, along with disorder defects which are defined by specifying a certain singular behavior as the gauge field approaches the defect. These surface defects are shown to engineer integrable field theories such as the Thirring model.

\end{itemize}

\paragraph{} What the aforementioned defects all have in common is that they are sitting at a fixed location in the holomorphic plane. On the other hand, it is natural to wonder about defects supported along the holomorphic plane and a fixed location in the topological plane. The most natural question to pose would be whether four-dimensional Chern-Simons theory can be consistently coupled to a two-dimensional \textit{holomorphic} field theory with global $G$ symmetry. Because the insertion of such a holomorphic surface defect, if it exists, still leaves room for the insertion of defects of various sorts along the topological surface, one can expect (with a healthy dose of optimism) that they will be related to universal phenomena in two-dimensional integrability.

\paragraph{} The purpose of the present paper is to establish the existence of holomorphic surface defects in four-dimensional Chern-Simons theory. The role of these defects in two-dimensional integrability will be discussed in a subsequent paper \cite{CIKY}.  

\paragraph{} The outline of this paper is as follows. In Section \ref{holft} we introduce the holomorphic sigma model and discuss the obstruction to coupling such a two-dimensional field theory to four-dimensional Chern-Simons theory. In Section \ref{framinganomaly} we derive the framing anomaly of four-dimensional Chern-Simons theory and show that if the curvature of the topological surface is sharply localized around the insertion point of the defect, the obstruction can be canceled. In Section \ref{monodromy}, we show as a corollary of this anomaly cancellation result, that four-dimensional Chern-Simons theory admits monodromy defects. In Section 5 we make some concluding remarks.

\section*{Acknowledgements} This paper would not have been possible without the generous guidance of Kevin Costello. I also thank Nafiz Ishtiaque for participating in the initial stages of this project and for many useful discussions. I would also like to acknowledge Kevin Costello, Nafiz Ishtiaque and Junya Yagi for collaboration on the upcoming sequel to this paper, and Edward Witten for many helpful discussions. This research is supported by the National Science Foundation grant PHY-1911298.

\section{Holomorphic Field Theories with Global Symmetry} \label{holft} \subsection{Classical Aspects} Let $\mathbb{C}$ denote the complex plane with the standard coordinates $(z, \bar{z})$.  A holomorphic field theory on $\mathbb{C}$ refers to a theory such that the purely antiholomorphic translations \bea z \rightarrow z, \,\,\,\,\,\,\,\, \bar{z} \rightarrow \bar{z} + \epsilon \eea are trivial. The \textit{holomorphic sigma model} for us will serve the dual purpose of being a canonical example that illustrates this notion, and being the actual defect theory we are interested in.

\paragraph{} Let $(X, I, \Omega)$ be a holomorphic symplectic manifold. This means that $X$ is a smooth manifold equipped with an integrable complex structure $I$, and a non-degenerate, $I$-holomorphic, closed two-form $\Omega$ (in particular the real dimension of $X$ is a multiple of $4$). $\Omega$ being a closed $2$-form $\del \Omega = 0$ means that locally we can write $\Omega = \del \Lambda$ for some (local) holomorphic one-form $\Lambda$ (as usual $\del$ denotes the $I$-holomorphic Dolbeault differential).  The basic field of the holomorphic sigma model consists of a map $$\phi: \mathbb{C} \rightarrow X.$$ The action of the theory is \bea \label{holaction} S = \frac{1}{2\pi}\int_{\mathbb{C}} \text{d}z \wedge \phi^*(\Lambda). \eea Letting $\big(\phi^i \big)_{i=1, \dots, \text{dim}_{\mathbb{C}}X}$ denote local $I$-holomorphic complex coordinates on $X$, the action in terms of these local coordinates reads \bea S = \frac{1}{2\pi}\int \text{d}^2 z \big( \Lambda_i \, \del_{\bar{z}} \phi^i \big).\eea The action $S$ is not real. Instead, it is to be thought of as a holomorphic functional on the complex field space $\text{Map}(\mathbb{C}, X)$. 

\paragraph{} The equation of motion for the holomorphic sigma model is \bea \Omega_{ij}\del_{\bar{z}} \phi^j = 0,\eea and because $\Omega$ is non-degenerate, it simply says that the map $\phi: \mathbb{C} \rightarrow X$ is holomorphic \bea \del_{\bar{z}} \phi^i = 0.\eea 

\paragraph{Remark} In order to get the holomorphic map equation as the equation of motion for an action functional, it is crucial that the target space be even complex dimensional (we had to invert $\Omega$). The standard Cauchy-Riemann equations in one complex dimension are not the variational equations of an action functional \cite{munoz1, munoz2}. 

\paragraph{} A convenient way of rewriting the action of the holomorphic sigma model is given by integrating \eqref{holaction} by parts: \bea S = -\frac{1}{2\pi}\int_{\mathbb{C}}\ z \,\phi^*(\Omega).\eea Writing the action this way makes it clear that it is independent of the choice of local Liouville one-form, and that it is single-valued.

\paragraph{} One can readily check that the current corresponding to the infinitesimal anti-holomorphic translation \bea \delta z = 0, \,\,\,\, \delta \bar{z} = \epsilon \eea simply vanishes \bea \delta_{\epsilon } S = 0\eea modulo total derivatives. Thus the $\bar{z}$-translations are trivial, and the theory is holomorphic. The holomorphic translations \bea \delta z = \epsilon, \,\,\,\,\,\,\,\, \delta \bar{z} = 0, \eea on the other hand lead to a holomorphic current $\del_{\bar{z}} T = 0$ where \bea T = \Lambda_i \del_z \phi^i.  \eea It is also natural to wonder if the holomorphic sigma model is invariant under the scaling transformation $$z \rightarrow \lambda  z, \,\,\,\,\,\,\,\, \bar{z} \rightarrow \bar{\lambda}\, \bar{z},$$ where $\lambda \in \mathbb{C}^{*}.$ The holomorphic sigma model for generic target $(X, \Omega)$ clearly does not have such a symmetry: the one form $\text{d}z$ scales but the Liouville one-form generically does not. However, it can be restored if $X$ admits a $\mathbb{C}^*$ action generated by a vector field $U$ under which the holomorphic symplectic form $\Omega$ transforms homogeneously \bea \mathcal{L}_{U} \Omega = \text{i} \alpha \Omega \eea for some weight $\alpha$. Then the infinitesimal transformation \bea \delta \phi^i = \epsilon( \frac{1}{\alpha} U^i + \text{i} z\del_z \phi^i - \text{i} \bar{z} \del_{\bar{z}} \phi^i)\eea becomes a symmetry of the theory. We will often assume the existence of such a $\mathbb{C}^*$-scaling action on $(X, \Omega)$ in this paper. 

\paragraph{} A special case of the holomorphic sigma model that is of particular interest is when $X$ is the holomorphic cotangent bundle of a complex manifold $Y$, $X = T^*Y$. Letting $\big(\gamma^A \big)$ be local coordinates on $Y$ and $\beta_A$ be the coordinates in the fiber direction, the holomorphic symplectic form on $X$ is \bea \Omega = \text{d}\beta_A \wedge \text{d} \gamma^A,\eea and so the Liouville one-form takes the standard form \bea \Lambda = \beta_A \text{d} \gamma^A,\eea and the action reads \bea S =  \frac{1}{2\pi}\int \text{d}^2 z \, \beta_A \del_{\bar{z}} \gamma^A. \eea The target space $\mathbb{C}^*$ transformation $$(\beta_A, \gamma^A) \rightarrow (\lambda \beta_A, \gamma^A)$$ gives $\Omega$ weight $\alpha = 1$ so that $\beta_A$ becomes a spin one field whereas $\gamma^A$ remains a scalar under $U(1)$ rotations. The holomorphic sigma model into a cotangent bundle $T^*Y$ is therefore the same as the non-linear $\beta\gamma$ system on $Y$. 

\paragraph{} Finally, it is important for us to discuss global symmetries in the holomorphic sigma model. Let $G$ be a complex Lie group with corresponding Lie algebra $\mathfrak{g}$. A holomorphic sigma model with target space $(X, \Omega)$ is said to have $G$-symmetry if the underlying target space has a subgroup inside the group of holomorphic symplectomorphisms that is isomorphic to $G$. Infinitesimally, this means that there is a Lie (sub)algebra of vector fields $\text{Symp}(X) \subset \Gamma(TX)$ defined by the condition $V \in \text{Symp}(X)$ if $$\mathcal{L}_V \Omega = 0.$$ $\text{Symp}(X)$ is a Lie subalgebra of $\Gamma(TX)$ with respect to the Lie bracket of vector fields. We then require that there is an injective Lie algebra homomorphism $\psi : \mathfrak{g} \rightarrow \text{Symp}(X)$. Choosing a basis of $\{t_a \}$ of $\mathfrak{g}$ such that the structure constants are $f^a_{bc}$, this simply means that there are $I$-holomorphic vector fields $K_a := \psi(t_a)$ such that \bea [K_a, K_b] = f_{ab}^c K_c\eea and $\mathcal{L}_{K_a} \Omega = 0$ for each vector field $K_a$. Letting $\mu: X \rightarrow \mathfrak{g}^{\vee}$ be the corresponding moment map, which by definition satisfies \bea \del_i \mu_a = K^j_a\Omega_{ji},  \eea the action under $\delta \phi^i = \epsilon^a K^i_a $ varies as \bea \label{variationundersymplecto} 2\pi \delta S = -\int \text{d}^2 z \,\del_{\bar{z}} \epsilon^a \mu_a.\eea Thus the holomorphic current corresponding to the $\mathfrak{g}$-symmetry is simply given by \bea J = -\mu. \eea In deriving this, we used that if $K_a$ generates a holomorphic symplectomorphism, the Liouville one-form varies as a total derivative: there is a local $\mathfrak{g}^{\vee}$-valued function $f$ such that \bea \mathcal{L}_{K_a} \Lambda = \text{d}f_a.\eea The moment map can then be expressed as \bea \mu_a = f_a - \iota_{K_a} \Lambda.\eea Classically, the $G$ symmetry can be coupled to a gauge field $A^a_{\bar{z}} \text{d}\bar{z}$ by adding the term \bea S_A = \frac{1}{2\pi}\int \text{d}^2 z \, A^a_{\bar{z}}  \mu_a\eea to the classical action. The total action \bea \label{gaugeaction} S = \frac{1}{2\pi}\int \text{d}^2 z \big( \Lambda_i \del_{\bar{z}} \phi^i + A_{\bar{z}}^a \mu_a \big) \eea is now invariant under the gauge transformations \bea \delta \phi^i &=& \varepsilon^a K_a^i , \\ \delta A^a_{\bar{z}} &=& \del_{\bar{z}} \varepsilon^a + f^{a}_{bc} A_{\bar{z}}^b \varepsilon^c. \eea The gauge invariance follows from how the first term varies \eqref{variationundersymplecto}, along with the fact that \bea K^i_a \del_i \mu_b = \{\mu_a, \mu_b \} = f_{ab}^c \mu_c,\eea where $\{ \, , \}$ denotes the Poisson bracket \bea \{f,g \} = \Omega^{ij}\del_i f \, \del_j g. \eea The equations of motion for the gauged sigma model are \bea \del_{\bar{z}} \phi^i - A_{\bar{z}}^a K^i_{a} &=& 0, \\ \mu_a &=& 0.\eea Here are three examples of theories with $G$-symmetry which are useful and illustrative.

\paragraph{Example: The Free $\beta\gamma$ System:} Let $R$ be a representation of $\mathfrak{g}$ with representation matrices $\rho(t_a): R \rightarrow R$. We can readily produce a free holomorphic field theory with $G$ symmetry, by considering the holomorphic sigma model into the linear space $X = T^*R$. The target space as a representation of $\mathfrak{g}$ is then simply $X = R \oplus R^{\vee}$. The $\mathfrak{g}$-currents/moment maps are given by \bea \mu_a = \beta_A  \, \rho(t_a)^A_{\,\,B} \gamma^B. \eea

\paragraph{Example: The Cotangent Bundle of a Flag Variety:} Another example that we will discuss extensively later is when $X = T^*Y$ where $Y$ is the flag manifold $G/B$ (here $B$ denotes the Borel subgroup of $G$). The underlying manifold $G/B$ has a $G$-action generated by holomorphic vector fields $K_a^A$ which can be naturally lifted to vector fields on $T^*(G/B)$ that preserve the canonical symplectic structure. The moment maps are \bea \mu_a = \beta_A K^{A}_a. \eea An important special case is $G= \text{SL}(2, \mathbb{C})$ where the corresponding flag variety $Y$ is the projective line $\mathbb{P}^1$. Letting $\gamma$ be a local coordinate on $\mathbb{P}^1$ (i.e a coordinate on one of the two standard patches), the $\mathfrak{sl}_2$ vector fields on $\mathbb{P}^1$ are \bea K^e = \frac{\del}{\del \gamma}, \,\,\,\,\,\,\, K^h = -2\gamma\frac{\del}{\del \gamma}, \,\,\,\,\,\,\, K^f = -\gamma^2 \frac{\del}{\del \gamma} .\eea Letting $\beta$ be the coordinate in the fiber direction, the corresponding moment maps then read \bea \mu^e &=& \beta, \\ \mu^h &=& -2\beta \gamma, \\ \mu^f &=& -\beta \gamma^2. \eea

\paragraph{Example: 4d Chern-Simons} As our third and final example, we note that four-dimensional Chern-Simons theory on $\Sigma \times \mathbb{C}$ can be considered as an example of a gauged holomorphic sigma model, where both the target space $X$ and the group $\mathcal{G}$ are infinite-dimensional\footnote{We have changed notation for the symmetry group $G$ to $\mathcal{G}$ for this particular example. $G$ will denote the gauge group of the Chern-Simons theory}. Let $\Sigma$ be a topological surface and $G$ a complex Lie group. The target space $X$ is given by the space of $G$-connections on $\Sigma$, \bea X = \{ G\text{-connections on } \Sigma \}  \eea which inherits a complex structure from the complex structure on $G$. Letting $x^{\alpha}$ denote local coordinates on $\Sigma$, the two-form \bea \label{sympform4dcs} \Omega = \int \text{d}^2 x \, \epsilon^{\alpha \beta} \text{Tr} \big(\delta A_\alpha \wedge \delta A_\beta \big),\eea where $\text{Tr}$ denotes an invariant bilinear form on $\mathfrak{g}$, provides a holomorphic symplectic structure on $X$. The group \bea \mathcal{G} = \text{Map}(\Sigma, G) \eea which acts infinitesimally on $X$ via gauge transformations \bea \delta_{\varepsilon} A_{\alpha} = \del_{\alpha} \varepsilon + [A_{\alpha}, \varepsilon],\eea provides us with holomorphic symplectomorphisms of $(X, \Omega)$. The corresponding moment map is \bea \mu(\varepsilon) = -\int \text{d}^2 x \, \epsilon^{\alpha \beta}\text{Tr}(\varepsilon F_{\alpha \beta}) \eea where $F$ denotes the curvature of the gauge field $A$ on $\Sigma$. Plugging in the local Liouville one-form \bea \Lambda = \int \text{d}^2 x \,\,\epsilon^{\alpha \beta} \text{Tr}(A_\alpha \delta A_\beta) \eea into the gauged sigma model action \eqref{gaugeaction}, we find \bea S = \frac{1}{2\pi}\int \text{d}^2 z \text{d}^2 x \big( \epsilon^{\alpha \beta}\text{Tr}( A_\alpha \del_{\bar{z}} A_\beta - A_{\bar{z}} F_{\alpha \beta} )\big). \eea This is precisely the action of four-dimensional Chern-Simons theory \bea S = \frac{1}{2\pi}\int_{\Sigma \times \mathbb{C}} \text{d}z \wedge \text{CS}(A) \eea on $\Sigma \times \mathbb{C}$ where the partial connection \bea A = A_{\alpha} \text{d}x^{\alpha} + A_{\bar{z}} \text{d}\bar{z}\eea is the fundamental field.

\paragraph{Generalizations} We now briefly mention some important generalizations of the holomorphic sigma model. The first generalization simply involves formulating the theory on a general Riemann surface $C$ once a closed holomorphic one-form $\omega$ has been chosen. The action is then simply \bea \label{actiononsurface} S = \frac{1}{2\pi} \int_{C} \omega \wedge \phi^*(\Lambda).\eea There is a further generalization which involves a \textit{family} of holomorphic symplectic forms on $X$ parametrized by the surface $C$. For this generalization, let $C$ be a Riemann surface with complex structure $j$, and let $(X,I)$ be a complex manifold. Suppose $\mathbb{T}$ is a closed, holomorphic $(3,0)$ form on the product $C \times X$ (equipped with the natural complex structure $j\oplus I$) which is \textit{vertically non-degenerate} \cite{munoz2}. Let $\mathbb{J}$ denote a local primitive $\text{d} \mathbb{J} = \mathbb{T}$ so that $\mathbb{J}$ is a $(2,0)$ form on $C \times X$.  We can write down a natural action as follows. For a map $\phi: C \rightarrow X$, let \bea \widetilde{\phi}: C \rightarrow C \times X \eea be the natural map defined via $\widetilde{\phi}(z) = \big(z, \phi(z)\big)$. We then define \bea S = -\frac{1}{2\pi}\int_{C} \widetilde{\phi}^*\big(\mathbb{J} \big).\eea When the three-form $\mathbb{T}$ is \bea \mathbb{T} = \omega \wedge \Omega \eea for a holomorphic one-form $\omega$ on $C$ and a holomorphic symplectic form $\Omega$ on $X$, we recover the action \eqref{actiononsurface}. 

\paragraph{Remark} The holomorphic sigma model into $(X, \Omega)$ has a holomorphic action functional, and so defining it non-perturbatively requires a choice of integration cycle in the field space $\text{Map}(\mathbb{C},X)$. Doing this via the gradient flow prescription described in \cite{Witten:2010zr}, one lands at the three-dimensional A-model with the same target space $(X, \Omega)$. Thus the three-dimensional A-model and the two-dimensional holomorphic sigma model have the same relationship as the two-dimensional A-model and analytically continued quantum mechanics \cite{Witten:2010zr}, and four-dimensional $\mathcal{N}=4$ Yang-Mills theory and analytically continued (three-dimensional) Chern-Simons theory \cite{Witten:2010cx}. The simplest instance of this relationship is that when $X=\mathbb{C}^2$ with standard symplectic form, the analytically continued theory is the A-twist of the three-dimensional $\mathcal{N}=4$ hypermultiplet \cite{Gaiotto:2016wcv}.

\subsection{Quantum Mechanical Considerations}

\paragraph{} Our discussion of holomorphic field theories and their global symmetries has been entirely classical so far. Working quantum mechanically requires more discussion. Quantum mechanically, holomorphicity of a field theory (along with $U(1)$ invariance) implies that the algebra of local observables is a vertex algebra. For an extensive discussion on this point, see \cite{Costello:2016vjw}, Chapter 5. One therefore expects to be able to carry out the quantum mechanical discussion entirely in the language of vertex algebras. We refer the reader to the review article \cite{kacnotes} for the basic formalism.

\paragraph{} If the holomorphic sigma model with target $(X, \Omega)$ exists at the quantum level, we should be able construct a well-defined vertex algebra $V(X, \Omega)$ \bea (X, \Omega) \leadsto V(X, \Omega).\eea However, this is not always possible; there can be obstructions to its existence. The obstructions are well-illustrated (and best understood) when $X$ is a cotangent bundle $T^*Y$ for a complex manifold $Y$ so that the theory in question is a non-linear $\beta\gamma$-system on $Y$. Here, the construction of a vertex algebra proceeds by covering $X$ by open sets $\{U_{\alpha}\}$. In each open set, the theory looks like a free $\beta\gamma$ system with singular operator product expansion \bea \gamma^A(z) \beta_B(w) \sim \frac{\delta^{A}_{\,\,B}}{z-w}.\eea We can then attempt to glue the different $\beta\gamma$-systems on overlaps $U_{\alpha }\cap U_{\beta}$ via an appropriate gluing rule and ask if our gluing laws are consistent. The obstruction to a consistent gluing rule is that the cohomology class of a degree $2$ cocycle valued in the sheaf of closed holomorphic $2$-forms on $Y$ vanishes. This cocycle can be shown to be equivalent to the first Pontryagin class $p_1(Y)$. If this is trivial, one can expect to get a well-defined vertex algebra $V(T^*Y)$. If it is non-trivial there can be no such expectation, and we say that the theory has a target space diffeomorphism anomaly. This was first derived in \cite{gerbepaper} and is reviewed in \cite{Witten:2005px} and \cite{Nekrasov:2005wg} from a more physical viewpoint \footnote{These papers also analyzed the obstructions to having a well-defined stress tensor. Namely, provided the $p_1(Y)$-anomaly vanishes and there is a well-defined vertex algebra, the obstruction to the vertex algebra having a conformal vector. This obstruction is measured by the first Chern class $c_1(Y)$. For us this will not play a role as we are not concerned with conformal invariance. Indeed, for most of our examples, this will be non-zero.}.

\paragraph{} It is interesting to generalize the obstruction theory applicable to cotangent bundles to arbitrary holomorphic symplectic manifolds with a $\mathbb{C}^*$-scaling symmetry. One of the main class of examples that will be discussed later is when $X$ is the coadjoint orbit of some element $\alpha \in \mathfrak{g}^{\vee}$ under the conjugation action of a complex Lie group $G$. Although for regular, semisimple $\alpha$ the space $X_{\alpha}$ is indeed not a cotangent bundle, the obstruction theory is nonetheless well-understood for this class of examples\footnote{As we will discuss later in the paper, coadjoint orbits are affine deformations of cotangent bundles.}. We will therefore not pursue the general obstruction theory in this paper. 

\paragraph{} Suppose that there are no obstructions to having consistent gluing rules across patches, so that there is a well-defined vertex algebra $$(V, T, \, |0 \rangle, Y(\cdot, z))$$ associated to $(X, \Omega)$. We are now interested in the quantum mechanical counterpart of $(X, \Omega)$ having an infinitesimal symplectomorphism algebra. The natural notion is as follows. 

\paragraph{} Recall that associated to a Lie algebra $\mathfrak{g}$ and a complex number $k$, there is a vertex algebra known as the affine vertex algebra $V_{k}\big(\widehat{\mathfrak{g}} \big)$. It is the vacuum module of the affine Kac-Moody algebra of $\mathfrak{g}$ \bea [t_a(n), t_b(m)] = f_{ab}^c t_c(n+m) + n k \delta(n+m,0) \kappa_{ab} \eea at level $k$.  We say that $V$ carries an affine $\mathfrak{g}$-symmetry at level $k$ if there is a vertex algebra homomorphism \bea \psi: V_{k}\big(\widehat{\mathfrak{g}} \big) \rightarrow V(X, \Omega).\eea Less formally stated, the quantum theory is required to have currents $J_a$ that satisfy the familiar current algebra operator product expansion \bea J_a(z) J_b(w) \sim \frac{k \kappa_{ab}}{(z-w)^2} + \frac{f_{ab}^{c} J_c(w)}{(z-w)}. \eea

\paragraph{} Going back to our standard examples, the vertex algebras and $\mathfrak{g}$-currents are as follows. Things are simplest for the free $\beta \gamma$ system in a representation $R$. Because of the linear nature of the target space, there is no gluing is required. The vertex algebra is several copies of the $\beta\gamma$ vertex algebra so that \bea \beta_A(z) \gamma^B(w) \sim -\frac{\delta_A^{\,\,B}}{(z-w)}.\eea The $\mathfrak{g}$-currents are given by \bea J_a(z) = \big(\beta_A \rho(t_a)^A_{\,\,B} \gamma^B \big)(z)\eea where $\big( \, \, \big)$ denotes the normally ordered product. One can compute that the level for these currents is given by $k$ where \bea \text{Tr}_R(\rho(t_a)  \rho(t_b)) = -k \delta_{ab}. \eea In particular for $R$ being the adjoint representation, we have $k = -2h^{\vee}$. 

\paragraph{} For the cotangent bundle to the flag variety the discussion is more involved. We discuss the case of $G=SL(2,\mathbb{C})$ in detail. The space $X= T^* \mathbb{P}^1$ is covered by two patches $U$ and $V$, and in each patch we have a free $\beta\gamma$-system, which are glued together on overlaps by the transformation rule \bea \label{gluegamma} \gamma' &=& \frac{1}{\gamma}, \\ \label{gluebeta} \beta' &=& -(\beta(\gamma \gamma) ) + 2 \del \gamma ,\eea where $(ab)$ denotes the normally ordered product of two fields $a$ and $b$. One can indeed work out that in each patch we find the expected singularity when taking the operator product of $\beta$ and $\gamma$, so that there is a consistent gluing law, and therefore no obstructions. The notion of $\mathfrak{sl}_2$ symmetry also carries over to the vertex algebra. Consider the currents \bea \label{p1e} e(z) &=& \beta(z), \\ \label{p1h} h(z) &=& 2(\gamma \beta)(z), \\ \label{p1f} f(z) &=& 2 \del \gamma (z)- (\beta(\gamma \gamma))(z) \eea written in the patch $V$. It can be shown that these indeed define globally well-defined currents across not just a patch but the entire space $X$, and that they satisfy the Kac-Moody $\mathfrak{sl}_2$ algebra at the critical level \bea k = -2 .\eea As is well-known, the Sugawara stress tensor \bea T = \frac{1}{2(k+ 2)} \big( \frac{1}{2} ( hh) + (ef) + (fe) \big)\eea ceases to be well-defined at the critical level, and so the vertex algebra is not a conformal one (recall that the obstruction to having a conformal vector was $c_1(Y) = c_1(\mathbb{P}^1) \neq 0$. ). It is also a noteworthy feature that the rescaled Sugawara current \bea S = (k+2)T =  \frac{1}{4} (hh) + \frac{1}{2}(ef) + \frac{1}{2}(fe)\eea simply vanishes when we plug in the currents above \bea S = 0.\eea The vertex algebra $V$ associated to the sigma model with target $T^* \mathbb{P}^1$ in ghost number zero (namely the global sections) is in fact the vacuum module of the affine $\mathfrak{sl}_2$ algebra at level $-2$, modulo the ideal generated by the singular vector corresponding to the field $S$,  \cite{Malikov:1998dw}.

\paragraph{} More generally, $Y$ being the flag variety $G/B$ gives an example of a non-linear $\beta\gamma$ system that is unobstructed ($p_1(G/B) = 0)$. As shown in \cite{Malikov:1998dw, Arakawa:2009cb} the vertex algebra associated to the $T^*(G/B)$ sigma model has Kac-Moody symmetry at critical level \bea k=-h^{\vee},\eea and the global sections of the sheaf of vertex algebras is the irreducible $\widehat{\mathfrak{g}}_{-h^{\vee}}$-module obtained from the vacuum module and quotienting by the center.  This example will play an important role in later sections. 

\paragraph{} Given a vertex algebra $V$ with some Kac-Moody currents, suppose we couple the currents to some gauge field $A_{\bar{z}} \text{d} \bar{z}$ via the interaction term \bea \frac{1}{2\pi}\int \text{d}^2 z A^a_{\bar{z}} J^a. \eea In order to investigate the question of whether gauge invariance continues to hold at the quantum level, one studies the effect of a gauge transformation on the partition function as a functional of $A$. Namely we study \bea \mathcal{A} = \delta_{\varepsilon} \, \text{log} \,Z[A], \eea where $\varepsilon$ is the gauge transformation parameter. A standard computation shows that this does not vanish. Instead the anomaly is given by \bea \mathcal{A} = \frac{k}{2\pi} \int \text{d}^2 z  \, \kappa_{ab} \, \big(\varepsilon^a \del_{z} A_{\bar{z}}^b \big).\eea We will refer to this as the anomaly to gauging a Kac-Moody global symmetry. 

\paragraph{} Given a holomorphic field theory with $G$-symmetry, the coupling to four-dimensional Chern-Simons theory involves choosing a point $w_0$ in the topological plane $\Sigma$. Once such a point is chosen, the action of four-dimensional Chern-Simons theory coupled to a holomorphic sigma model into $(X, \Omega)$ is \bea S = \frac{1}{2\pi} \int_{\Sigma \times \mathbb{C}} \text{d}z \wedge \text{CS}(A) + \frac{1}{2\pi} \int_{\mathbb{C}} \text{d}z \wedge \big( \phi^*(\Lambda) + i^*\big(\mu(A) \big) \big),\eea where $i: \mathbb{C} \rightarrow \Sigma \times \mathbb{C}$ denotes the embedding $z \rightarrow (w_0, z)$. By assumption the classical moment maps $\mu_a$ have appropriate quantizations such that quantum mechanically they define currents satisfying the Kac-Moody algebra for some level $k$. Therefore the coupling of the bulk four-dimensional gauge field to these currents contributes to an anomaly. Unless there is some mechanism to cancel this, there is no gauge invariant way to couple a holomorphic sigma model with $G$-symmetry to four-dimensional Chern-Simons theory. The content of the next section is to show that the \textit{framing anomaly} of four-dimensional Chern-Simons provides us with such a mechanism. 

\section{The Framing Anomaly}  \label{framinganomaly} 

\subsection{The Framing Anomaly on $\Sigma \times \mathbb{C}$}

\paragraph{} Classical four-dimensional Chern-Simons theory is independent of any choice of metric on the surface $\Sigma$. From point of view of the holomorphic sigma model, this is because the holomorphic symplectic form \eqref{sympform4dcs} on the space of $G$-connections on $\Sigma$ is purely topological. We therefore say that four-dimensional Chern-Simons theory, classically, is topological along $\Sigma$. 

\paragraph{} At the quantum level, gauge invariance of the path integral requires one to make a choice of gauge fixing. The standard way of doing this is by introducing a metric on $\Sigma \times C$ and imposing an appropriate Lorentz gauge fixing condition with respect to this metric. It turns out that the quantum theory is not independent of the choice of metric that was made in defining it pertubatively. There is a mixed gravitational-gauge anomaly, analogous to the framing anomaly of three-dimensional Chern-Simons theory \cite{Witten:1988hf}. For a version of this anomaly applicable to Wilson lines, see \cite{Costello:2017dso}.

\paragraph{} Before embarking on the derivation of the framing anomaly, we set our conventions. The gauge algebra $\mathfrak{g}$ of the theory is taken to be a complex, simple Lie algebra with generators $\{t_a\}$ satisfying \bea [t_a, t_b] = f^{c}_{\,\,ab} t_c,\eea and is equipped with a Killing form $\kappa$, normalized so that the formula \bea \label{ffh} f_{abc}f^{bc}_{\,\,\,e} = -2h^{\vee} \kappa_{ae},\eea where $h^{\vee}$ denotes the dual Coxeter number of $\mathfrak{g}$, holds. The theory is formulated on the spacetime manifold \bea M = \Sigma \times \mathbb{C}\eea where $\Sigma$ is a topological surface and $\mathbb{C}$ is the complex plane. We choose $x,y$ to be local coordinates along $\Sigma$, and $z, \bar{z}$ to be the standard complex coordinates on $\mathbb{C}.$ As discussed in Section 2, the basic field of four-dimensional Chern-Simons theory is a $\mathfrak{g}$-valued partial connection on $M$ of the form \bea A = A_x \text{d}x + A_y \text{d}y + A_{\bar{z}} \text{d}\bar{z}.\eea Finally, we can write down the action of 4d Chern-Simons. It reads \bea S = \frac{1}{2\pi} \int \text{d}z \wedge \text{CS}(A)\eea where $\text{CS}(A)$ denotes the standard Chern-Simons three-form of a connection $A$ \bea \text{CS}(A) = \kappa\big(A,\text{d}A + \frac{2}{3}[A,A] \big).\eea In terms of explicit coordinates it reads \bea \label{action} S = \frac{1}{\pi} \int \text{d}^2x \text{d}^2 z \, \kappa_{ab} \big(A^a_y \del_{\bar{z}} A^b_x + A^a_{\bar{z}} \del_x A^b_y + A^a_x \del_y A^b_{\bar{z}} + f^b_{\,\,cd} A^a_{\bar{z}} A^c_x A^d_y \big).\eea Just like the holomorphic sigma model, $S$ is to be regarded as a holomorphic function on the space of partial connections on $\Sigma \times \mathbb{C}.$ 

\paragraph{} We now turn to a derivation of the framing anomaly of four-dimensional Chern-Simons theory. The equation of motion of the theory is \bea \text{d}z \wedge F = 0,\eea where $F = \text{d}A + A \wedge A$ denotes the curvature of the partial connection $A$. In individual components this says \bea F_{xy} = F_{x\bar{z}} = F_{y\bar{z}} = 0.\eea  We are interested in the quantum effective action $\Gamma[A]$ as a functional of a background field $A$ that solves these equations of motion. The framing anomaly comes about when we compute the variation of the effective action $\Gamma[A]$ at one-loop, under gauge transformations of the background field $A$: \bea \mathcal{A}[A] = \delta_{\varepsilon} \, \Gamma^{(1\text{-loop})}[A]. \eea

\paragraph{} Let us therefore formulate the quantum effective action $\Gamma[A]$ more precisely. The classical action when evaluated on \bea  A + B\eea where $A$ is a background solution to the equations of motion, and $B$ is a fluctuation is \bea S[A+B] = S[A] + \frac{1}{2\pi} \int \text{d}z \,\kappa\big(B , \text{d}_{A} B + \frac{2}{3}[B,B] \big)  \eea where $\text{d}_{A} B = \text{d}B + [A, B]$.  The action in the fluctuation field $B$ is invariant under the gauge transformation \bea B \rightarrow B+ \text{d}_{A} \epsilon + [B, \epsilon],\eea and therefore defining the path integral requires a choice of gauge fixing. An elegant way of doing this is by using the Batalin-Vilkovisky (BV) formalism \footnote{For the discussion of the BV complex we keep $C$ the holomorphic surface general.}. 

\paragraph{} The BV formalism involves the introduction of a ghost field $c$, an anti-field $B^{\vee}$ of the fluctuation field $B$, and an anti-field $c^{\vee}$ of the ghost field $c$. The full BV field space consisting of $(c, B, B^{\vee}, c^{\vee})$ is a differential graded Lie algebra with an odd symplectic pairing. For four-dimensional Chern-Simons theory on $\Sigma \times C$ this differential-graded Lie algebra can be nicely formulated in terms of $\mathfrak{g}$-valued differential forms on $\Sigma \times C$. The BV field space, which we will denote as $\Omega_{\text{CS}_4}(\Sigma \times C; \mathfrak{g})$ is given by \bea \Omega^*_{\text{CS}_4}(\Sigma \times C; \mathfrak{g}) = \bigoplus_{p+q= *} \Omega^p_{\text{dR}}(\Sigma; \mathfrak{g}) \otimes \Omega^{(0,q)}_{\bar{\del}}(C; \mathfrak{g}),  \eea where $\Omega^{p}_{\text{dR}}(\Sigma; \mathfrak{g})$ denotes the space of $\mathfrak{g}$-valued $p$-forms on $\Sigma$, and $\Omega^{(0,q)}_{\bar{\del}}(C; \mathfrak{g})$ denotes the space of $\mathfrak{g}$-valued $(0,q)$ forms on $C$. There is a gradation on the BV field space by the form degree $F$. It is related to the ghost number (i.e homological grading) by \bea \text{gh} = 1-F. \eea More explicitly, in ghost number $1$ there is the ghost field \bea c \,\, \in \,\, \Omega^0_{\text{CS}_4}(\Sigma \times C ; \mathfrak{g}),\eea in ghost number $0$, we have the one-form field \bea B_\alpha \text{d}x^{\alpha} + B_{\bar{z}} \text{d}\bar{z} \,\, \in \,\, \Omega^1_{\text{CS}_4} (\Sigma \times C; \mathfrak{g}),\eea in ghost number $-1$, we have the two form field \bea B^{\vee}_{\alpha \beta} \text{d}x^{\alpha} \wedge \text{d}x^{\beta} + B^{\vee}_{\alpha \bar{z}} \text{d}x^{\alpha} \wedge \text{d} \bar{z} \,\, \in \,\, \Omega^2_{\text{CS}_4} (\Sigma \times C; \mathfrak{g}), \eea and finally in ghost number $-2$, the field is a three-form field \bea c^{\vee}_{\alpha \beta \bar{z}} \text{d}x^{\alpha} \wedge \text{d}x^{\beta} \wedge \text{d}\bar{z} \,\, \in \,\, \Omega^3_{\text{CS}_4} (\Sigma \times C; \mathfrak{g}).\eea The Lie algebra structure is given by combining the wedge product of forms with the Lie bracket on $\mathfrak{g}$. The Lie bracket has ghost number $-1$. The differential on $\Omega^{*}_{\text{CS}_4}(\Sigma \times C; \mathfrak{g})$ is given by \bea \text{d}_{\text{CS}_4} = \text{d}_A(\Sigma) + \ov{\del}_A(C), \eea the sum of deRham gauge exterior derivative $\text{d}_A$ along $\Sigma$, and the Dolbeault exterior derivative $\ov{\del}_A$ along $C$. This is a nilpotent operator \bea \text{d}_{\text{CS}_4}^2 = 0 \eea because $A$ satisfies the equations of motion.  The odd symplectic pairing is given by \bea \alpha, \beta \in \Omega^{*}_{\text{CS}_4} (\Sigma \times C ; \mathfrak{g}) \rightarrow \langle \alpha, \beta \rangle := \int_{\Sigma \times C} \omega \wedge \kappa \big(\alpha, \beta \big) \,\,\in \,\, \mathbb{C}, \eea where as before $\omega$ is a closed holomorphic one-form on $C$. Letting $X \in \Omega^{*}_{\text{CS}_4}(\Sigma \times C; \mathfrak{g})$ be a field in the BV field space, the BV action is the generalized Chern-Simons action \bea S_{\text{BV}}[X] = \frac{1}{2\pi} \langle X, \text{d}_{\text{CS}_4} X + \frac{2}{3}[X,X]  \rangle. \eea It is useful to write it down in terms of the individual fields: \begin{align} \begin{split} S_{\text{BV}} = \frac{1}{2\pi} \int_{\Sigma \times C} \omega \wedge \kappa(B, \,\text{d}_A B + \frac{2}{3}[B,B]) + \frac{1}{\pi} \int_{\Sigma \times C} \omega \wedge \kappa ( B^{\vee}, \text{d}_A c + [B,c] ) \\ + \frac{1}{\pi} \int_{\Sigma \times C} \omega \wedge \kappa( c^{\vee},[c,c]). \end{split} 
\end{align} So far we have simply extended the field space while preserving the holomorphic-topological nature of the theory. We now have to choose a gauge fixing condition. In the BV formalism, this is done by choosing a Lagrangian subspace $L$ of the BV field space such that the quadratic part of the action becomes non-degenerate along $L$. A natural gauge fixing condition for four-dimensional Chern-Simons theory is as follows. We pick a Riemannian metric\footnote{More specifically, a Riemannian metric along $\Sigma$ and a K\"{a}hler metric along $C$.} $g_{\Sigma} \oplus g_C$ on $\Sigma \times C$ and define the operator \bea \delta_{A} =  \big(d_{A}(\Sigma) \big)^{\dagger} + 2 \big( \ov{\del}_A(C) \big)^{\dagger} \eea where $d_{A}(\Sigma)^{\dagger}$ is the natural adjoint of $d_A(\Sigma)$ on the space of differential forms on $\Sigma$, and $\ov{\del}_{A}^{\dagger }$ denotes the natural adjoint on $\ov{\del}$ on the space of anti-holomorphic forms on $C$ with respect to the metric $C$. Explicitly, these operators read \bea d_A(\Sigma)^{\dagger} &=& g^{\alpha \beta} \iota_{\frac{\del}{\del x^{\alpha}}} D_{\beta}, \\ \ov{\del}_A(C)^{\dagger} &=& g^{z\bar{z}} \iota_{\frac{\del}{\del \bar{z}}} D_z = g^{z\bar{z}} \iota_{\frac{\del}{\del \bar{z}}} \del_z, \eea where $D_{\alpha}$ denotes the covariant derivative involving both the background gauge field $A$, and the Christoffel connection on $\Sigma$. The covariant derivative $D_z$ when acting on anti-holomorphic forms on $C$ is simply the ordinary derivative $\del_z$ because there is no $z$-component of the connection $A$, and there are no mixed $z\bar{z}$ components in the Christoffel connection on $C$. In particular the action of the operator $\delta_A$ on one-forms $$\delta_A : \Omega^{1}_{\text{CS}_4} \rightarrow \Omega^{0}_{\text{CS}_4}$$  is given by \bea \delta_A(B_{\alpha} \text{d}x^{\alpha} + B_{\bar{z}} \text{d}\bar{z}) = g^{\alpha \beta} D_{\alpha} B_{\beta} + 2g^{z\bar{z}} \del_{z} B_{z}.\eea The main property of $\delta_A$ is that the operator \bea \Delta(A,g) := \delta_A d_A + d_A \delta_A, \eea when $A = 0$ and $\mathfrak{g}=\mathbb{C}$ becomes the standard Hodge Laplacian on $\Sigma \times C$ with respect to the product metric $g_{\Sigma} \oplus g_{C}$\footnote{That this be the case is why one has to introduce the factor of $2$ in the definition of $\delta_A$. One must remember that on a K\"{a}hler manifold the Dolbeault Laplacian and the Hodge Laplacian are related by \bea 2(\ov{\del} \, \ov{\del}^{\dagger} + \ov{\del}^{\dagger} \,\ov{\del}) = d^{\dagger}d + d d^{\dagger}. \eea} acting on the space $\Omega^{*}_{\text{dR}}(\Sigma) \otimes \Omega^{(0,*)}_{\ov{\del}}(C)$. The Lagrangian subspace $L$ is defined to be the kernel of $\delta_A$ \bea L = \{X \in \Omega^*_{\text{CS}_4} | \delta_A X = 0 \}.\eea In particular this imposes the Lorentz type gauge fixing condition on the one-form field $B$ that says \bea g^{\alpha \beta} D_{\alpha} B_{\beta} + 2g^{z\bar{z}} \del_{z} B_{z} = 0,\eea where $D_{\alpha}$ is the covariant derivative involving both the gauge field components along $\Sigma$, and the Christoffel connection on $\Sigma$.  The path integral \bea Z[A] = \int_{X \in L} \text{D}X \, e^{-S[X]}\eea then is formally non-degenerate\footnote{Here we are implicitly assuming that the BV complex $(\Omega^{*}_{\text{CS}_4}, d_{\text{CS}_4})$ has trivial cohomology, and that the holomorphic one-form has no zeros.}, and the quantum effective action is defined as its logarithm \bea \Gamma[A] = -\text{log}\, Z[A].  \eea

\paragraph{} Let's now come to the one-loop effective action which means we only keep the quadratic part of the BV action. Formally, this is given by the superdeterminant of the operator $\text{d}_A$ \bea Z^{(1\text{-loop})}[A] = \text{Sdet}\big(\text{d}_A : L \rightarrow L^{\perp} \big)^{-\frac{1}{2}},\eea where $\text{Sdet}$ denotes the superdeterminant with respect to the ghost number. By some standard manipulations, this can be further shown to be equivalent to the (square-root of the) ``holomorphic-topological" torsion defined as follows: Given the operator \bea \Delta = \delta_A \text{d}_A + \text{d}_A \delta_A \eea acting on $\Omega^*_{\text{CS}_4}$, the holomorphic-topological torsion on $\Sigma \times C$ with respect to the metric $g$ and background connection $A$ is defined entirely analogous fashion to the standard topological torsion \bea \text{log} \, \tau(\Sigma \times C, g, A) = -\frac{1}{2} \int_{\epsilon}^{\infty} \frac{\text{d}\tau }{\tau}\, \text{Tr}_{\,\Omega^*_{\text{CS}_4}(\Sigma \times C; \mathfrak{g})} \big(F(-1)^F e^{\tau \Delta} \big).  \eea A more precise way to regularize the divergences in this expression is by $\zeta$-function regularization as is standard in the literature on analytic torsion \cite{67624}. 

\begin{figure}
\centering
\includegraphics[width=0.92\textwidth]{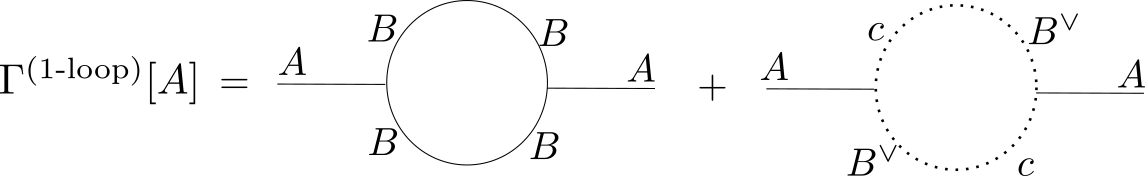}
\caption{Diagrams that contribute to the one-loop effective action as a functional of $A$.}  
\label{diagrams}
\end{figure}

\paragraph{} The point of the framing anomaly is that in the holomorphic-topological setting, the torsion is not an invariant of the gauge equivalence class of the background connection $A$, provided  that $\Sigma$ has non-trivial curvature. Its variation under gauge transformations is captured by the variation of the simple one-loop diagrams depicted in Figure \ref{diagrams}. 

\paragraph{} We now specialize to $C=\mathbb{C}$ again, picking the metric with $g_{z\bar{z}}= \frac{1}{2}$. 

\paragraph{} In order to compute the one-loop diagrams of Figure \ref{diagrams}, we need to know the propagator. The propagator, $P$ defined as the formal inverse of the operator $\frac{1}{\pi} \text{d}_{\text{CS}_4} : L \rightarrow L^{\perp}.$ It is given by \bea P(x,y) = \pi \delta_{A,x}G(x,y),\eea where $G$ is the Green function of the Hodge Laplacian acting on $L^{\perp} = \text{Ker}d_A \subset \Omega^*_{\text{CS}_4}$. $G$ is the integral kernel of the formal inverse of the Hodge Laplacian $\Delta$. The propagator is naturally a $\mathfrak{g} \otimes \mathfrak{g}^{\vee}$-valued two-form on $(\Sigma \times C)^2$, since $G$ is naturally a $\mathfrak{g}\otimes \mathfrak{g}^{\vee}$-valued three-form and $\delta_A$ reduces the form degree by one. In particular, the two-point function of the gauge field fluctuation $B$ is given by \bea P_{BB}(x,y) = \pi \delta_{A, x} G(x,y)|_{ \Omega^1_{\text{CS}_4}},\eea and the $c B^{\vee}$ propagator is given by \bea P_{c B^{\vee}}(x,y) = \pi \delta_{A, x} G(x,y)|_{\Omega^0_{\text{CS}_4}}. \eea  Let's work out the diagram in which the gauge fluctuation field $B$ propagates in the loop, and the diagram with the same underlying graph where the ghost and the anti-field to the gauge field fluctuation propagate in the loop, separately. As in \cite{Costello:2017dso} (and also \cite{Axelrod:1991vq}), perturbation theory is carried out conveniently in terms of differential forms. The propagator $P$ is a two-form, and the interaction vertex is the one-form \bea \frac{1}{2\pi} \text{d}z f_{abc}.\eea The amplitude corresponding to the $BB$ loop is given by \bea \frac{1}{2}\big(\frac{1}{2\pi} \big)^2 \int_{(\Sigma \times C)^2} f_{abc}  f_{def} \, \text{d}z_1 \, \text{d}z_2 \wedge A^a(x_1) \wedge P^{bc}_{BB}(x_1, x_2) \wedge P^{de}_{BB}(x_2, x_1)  \wedge A^f(x_2) .\eea The integrand is indeed an eight-form on $(\Sigma \times C)^2$. The amplitude is UV divergent. A convenient regularization involves uses the heat kernel. We write \bea G(x,y) = \int_0^{\infty} \text{d} \tau \, K_{\Omega^{*}_{\text{CS}_4}}(x,y ; \tau) \eea where $K_{\Omega^*_{\text{CS}_4}}(x,y; \tau)$ is the heat kernel of the Hodge Laplacian $\Delta(A,g)$ acting on $\Omega^{*}_{\text{CS}_4}(\Sigma \times \mathbb{C}; \mathfrak{g})$. More explicitly, the heat kernel is given as follows. Letting $\{\omega_{n} \}$ denote a basis of eigenforms of $\Delta(A,g)$ with corresponding eigenvalues $\{\lambda_n\}$. Then \bea K_{\Omega^*_{\text{CS}_4}}(x,y; \tau)= \sum_{n } e^{\lambda_n \tau} \big(*\omega_{n}(x) \big) \wedge \big(\omega_n(y) \big) \eea where the Hodge $*$ is defined by using the three-dimensional $\epsilon$-symbol with $\epsilon^{xy \bar{z}} = 1$. In particular, it maps a $p$-form in the BV field space to a $3-p$ form, and \bea *1 = \frac{\text{i}}{2}\sqrt{g_{\Sigma}} \, \text{d}x \wedge \text{d}y \wedge \text{d} \bar{z}. \eea We therefore see that the heat kernel is a $\mathfrak{g}^{\otimes 2}$-valued three-form on $(\Sigma \times C)^2$. The $BB$ propagator is then given by \bea P_{BB}(x,y) = \pi \int_0^{\infty} \text{d} \tau \, \delta_{A,x} K_{\Omega^{1}_{\text{CS}_4}}(x,y; \tau). \eea  We are interested in the amplitude to  quadratic order in the background gauge field $A$. Because there are already two external $A$-fields, we can use the heat kernel for $\Delta(A,g)$ with $A=0$. It takes the form \bea K^{ab}_{\Omega^{1}_{\text{CS}_4}(\Sigma \times C; \mathfrak{g})}(x,y; \tau)|_{A_0 = 0 }= \kappa^{ab} K_{\Omega^{1}_{\text{CS}_4}(\Sigma \times C)}(x,y; \tau) \eea  where the factored out part is the purely geometric heat kernel for the Hodge Laplacian acting on one-forms. We can now write the amplitude as \begin{align} \frac{1}{4}h^{\vee}\kappa_{ab} \int \text{d}\tau_1 \text{d} \tau_2 \text{d}z_1 \text{d}z_2 \,A^a(x_1) \wedge \delta_{x_1} K_{\Omega^1_{\text{CS}_4}(\Sigma \times C)}(x_1, x_2,\tau_1) \wedge \delta_{x_2} K_{\Omega^1_{\text{CS}_4}(\Sigma \times C)}(x_2, x_1,\tau_2) \wedge A^b(x_2). 
\end{align} In order to compute the anomaly of this amplitude, it is now convenient to specialize to a particular solution of the equation of motion. We consider $$A = A_{\bar{z}}(z, \bar{z}) \text{d}\bar{z},$$ which is a solution provided $A_{\bar{z}}(z, \bar{z})$ is constant along $\Sigma$. We now use the factorization property of the heat kernel, which says \begin{align}  K_{\Omega^p(M_1 \times M_2)}\big( (x_1, x_2), (y_1, y_2);\tau ) = \sum_{r+s = p} K_{\Omega^r(M_1)} \big( x_1, y_1 ;\tau \big) \wedge K_{\Omega^{s}( M_2)}\big( x_2, y_2 ;\tau \big).
 \end{align} Applied to $K_{\Omega^1_{\text{CS}_4}(\Sigma \times \mathbb{C})}$, we find that the heat kernel can be written as a sum of two terms; the first term being the heat kernel for zero-forms on $\Sigma$ times the heat kernel for anti-holomorphic one-forms on $\mathbb{C}$, and the other term with the form degrees reversed. The first term is a $(2,0)$ form along $\Sigma \times \Sigma$ and a $(0,1)$ form along $\mathbb{C}$, so when $\delta$ acts on the first factor, it gives a form of mixed degree. On the other hand, the second term is a $(1,1)$ form along $\Sigma$ and a $(1,0)$ form along $\mathbb{C} \times \mathbb{C}$. When $\delta$ acts one the second factor, it results in a $(1,1)$ form along $\Sigma \times \Sigma$. Remembering that $g_{z\bar{z}} = \frac{1}{2}$ so that $$2\ov{\del}^{\dagger} = 4 \del_z,$$ the amplitude becomes \begin{align} \begin{split} 4h^{\vee} \kappa_{ab} \int \text{d}\tau_1 \text{d}\tau_2 \text{d}^2 z_1 \text{d}^2 z_2 A^a_{\bar{z}}(z_1,\bar{z_1}) A^b_{\bar{z}}(z_2, \bar{z_2}) K_{\Omega^1(\Sigma)}((x_1, y_1), (x_2, y_2); \tau_2 ) \,\,\,\,\,\,\,\,\,\, \\ \wedge K_{\Omega^1(\Sigma)}((x_2, y_2), (x_1, y_1); \tau_2 ) \del_{z_1} K_{\mathbb{C}}(z_1, z_2 ; \tau_1) \del_{z_2} K_{\mathbb{C}}(z_2, z_1 ;\tau_2) . \end{split}
 \end{align} We can now use the composition property of the heat kernel \bea \int_{M_y} K_{\Omega^*(M)}(x,y, \tau) \wedge  K_{\Omega^*(M)}(y,z ,\tau') = K_{\Omega^*(M)} (x,z, \tau + \tau') \eea to integrate along the copy of $\Sigma$ parametrized by $(x_2, y_2)$. This leaves us with an integrand involving only the diagonal form of the one-form heat kernel on $\Sigma$ \bea K_{\Omega^1(\Sigma)}(x,x; \tau).\eea The one-form heat kernel on a surface $\Sigma$ has the well-known \cite{McKean:1967xf} short-time asymptotics \bea  K_{\Omega^1(\Sigma)}(x,x; \tau) = \frac{\text{dvol}}{4\pi \tau} - \frac{R(x)  \text{dvol}}{12\pi} + O(\tau), \eea where $R$ denotes the Ricci scalar of $\Sigma$. Let us focus on the contribution of the $\tau$-independent term to the amplitude (the other terms will cancel against the ghost-antifield loop). We are left with \begin{align} -4h^{\vee} \times \frac{1}{12\pi} \int_{\Sigma} \text{dvol}_{\Sigma} \,R_{\Sigma}(x) \int \text{d}\tau_1 \text{d}\tau_2 \text{d}^2 z_1 \text{d}^2 z_2 A^a_{\bar{z}}(z_1, \bar{z}_1) A^b_{\bar{z}}(z_2, \bar{z}_2) \del_{z_1} K_{\mathbb{C}}(z_1, z_2, \tau_1) \del_{z_2}K_{\mathbb{C}}(z_2, z_1, \tau_2). 
 \end{align} We can now use the explicit form of the heat kernel on $\mathbb{C}$ \bea K_{\mathbb{C}}(z_1, z_2; \tau) = \frac{1}{4\pi \tau} e^{-\frac{|z_1-z_2|^2}{4\tau}} \eea and perform the $(\tau_1, \tau_2)$ integrals to give \bea 4h^{\vee} \times \frac{1}{12\pi} \times \Big(\frac{1}{4\pi} \Big)^2 \int_{\Sigma} \text{dvol}_{\Sigma} \,R_{\Sigma}(x) \text{d}^2 z_1 \text{d}^2 z_2 A^a_{\bar{z}}(z_1, \bar{z}_1) A^b_{\bar{z}}(z_2, \bar{z}_2) \frac{1}{(z_1-z_2)^2}. \eea Upon performing a gauge transformation $A_{\bar{z}} \rightarrow D_{\bar{z}} \varepsilon$ and keeping the linear term in $A$, we find by using $$\del_{z_1} \frac{1}{(z_1-z_2)^2} = -2\pi \del_{\bar{z}_1}\delta^{(2)}(z_1-z_2, \bar{z}_1 - \bar{z}_2)$$ that \bea \mathcal{A}_{BB} = \frac{h^{\vee}}{12\pi^2} \int \text{d}\text{vol}_{\Sigma} \, \text{d}^2 z \, R_{\Sigma} \text{Tr}(\varepsilon\, \del_z A_{\bar{z}}).  \eea The calculation of the anomaly of the diagram where the ghost field and gauge anti-field propagate is entirely analogous. The main point there is that the calculation involves the short-time asymptotics of the diagonal heat kernel acting on zero-forms on $\Sigma$ instead of one-forms. These are well-known to be \bea K_{\Omega^0(\Sigma)}(x,x; \tau) = \frac{\text{dvol}}{4\pi \tau} + \frac{R(x) \text{dvol}} {24 \pi} + O(\tau). \eea Note that the sign of the $\tau$-independent term in the asymptotic expansion is crucially flipped when compared to the diagonal one-form heat kernel. On the other hand, the Grassman-odd nature of the fields that propagate in this loop, lead to another minus sign. The result is that the anomaly coming from the $cB^{\vee}$ loop is $\frac{1}{2}$ times the $BB$-anomaly \bea \mathcal{A}_{cB^{\vee}} = \frac{h^{\vee}}{24\pi^2} \int \text{d}\text{vol}_{\Sigma} \, \text{d}^2 z \, R_{\Sigma} \text{Tr}(\varepsilon\, \del_z A_{\bar{z}}). \eea Note that in particular, the singular term in $\tau$ cancels when adding up the $BB$ contribution and the $cB^{\vee}$ contributions. The total anomaly $$\mathcal{A} = \mathcal{A}_{BB} + \mathcal{A}_{cB^{\vee}}$$ is therefore given by \bea \label{anomalyresult} \mathcal{A} = \frac{h^{\vee}}{8\pi^2} \int_{\Sigma \times \mathbb{C}} \text{d}\text{vol}_{\Sigma} \, \text{d}^2 z \, R_{\Sigma} \text{Tr}(\varepsilon \,\del_z A_{\bar{z}}). \eea This is the final form of the framing anomaly. 
 
\paragraph{Remark: A Shortcut to the Proportionality Factor} We can verify the prefactor $\frac{h^{\vee}}{8\pi^2}$ in a quick way by using the following argument due to K.~Costello \cite{Cpriv}. Suppose we specialize to $\Sigma = S^2$ and choose a metric with radius $R$ and send $R \rightarrow 0$. In this limit, only the harmonic forms on $S^2$ survive. There are no harmonic one-forms, so the fluctuation field $B$ along the $S^2$ direction simply vanishes. $H^0(S^2)$ and $H^2(S^2)$ are each one-dimensional, and so there are modes of the ghost and anti-field that survive. The quadratic BV action is now the action of a gauged, adjoint $bc$ ghost system \bea S_{bc} =  \frac{1}{2\pi} \int_{\mathbb{C}}  b \,\ov{\del}_A c,\eea where $c$ is a $0$-form (coming from the original ghost field $c$) and $b$ is a $(1,0)$ form on $\mathbb{C}$ (coming from $\text{d}z \wedge B^{\vee}$). Under a gauge transformation, this has an anomaly of the form\footnote{The adjoint-valued $\beta\gamma$ system has $k=-2h^{\vee}$ and so the proportionality factor of its Kac-Moody anomaly is $\frac{k}{2\pi} = -\frac{h^{\vee}}{\pi}$. The anomaly for the $bc$ ghost system can be obtained by an overall sign flip.} \bea \mathcal{A}_{bc} = \frac{h^{\vee}}{\pi} \int \text{d}^2 z \,\text{Tr} \big(\varepsilon \, \del_z A_{\bar{z}} \big).\eea On the other hand, if the anomaly is of the form $c \int \text{dvol}_{\Sigma} R_{\Sigma}  \text{d}^2 z \text{Tr}\big( \varepsilon \del_z A_{\bar{z}} \big) $ for some constant $c$, upon integrating along the $S^2$ direction and equating with the $bc$ ghost anomaly, we get \bea 4\pi c \, \chi(S^2) = \frac{h^{\vee}}{\pi}.\eea We conclude that \bea c = \frac{h^{\vee}}{8\pi^2}.\eea

\paragraph{} Before going on to discuss the anomaly cancellation result for holomorphic surface defects, we pause for a bit and discuss what the above result means for four-dimensional Chern-Simons theory without any defects. The result \eqref{anomalyresult}, means that four-dimensional Chern-Simons theory on $\Sigma \times \mathbb{C}$ suffers from a gauge anomaly unless the curvature of $\Sigma$ vanishes. This means that we must require that $\Sigma$ have a trivial tangent bundle. Suppose then that $\Sigma$ is a parallelizable surface, and moreover we pick a trivialization. Picking a trivialization amounts to picking a trivializing one-form $\rho$ for the Euler class of $\Sigma$. What happens under changing the trivializing one-form $\rho$ by a total derivative $\rho \rightarrow \rho + \text{d}f$ for some function $f$ on $\Sigma$? The answer turns out to be that the effective action has an anomaly of the form \bea \label{framingdep} \mathcal{A}_f = \frac{\alpha h^{\vee}}{2\pi} \int_{\Sigma \times \mathbb{C}} f \,\text{Tr} \, F \wedge F,\eea for some dimensionless constant $\alpha$. Thus we find that not only is $\Sigma$ required to have a trivial tangent bundle, but four-dimensional Chern-Simons theory detects the choice of trivialization. We therefore require $\Sigma$ be a \underline{framed} surface. 

\paragraph{} Note that by writing the four-dimensional Chern-Simons action as \bea S = -\frac{1}{2\pi} \int z \text{Tr}(F \wedge F),\eea the anomaly \eqref{framingdep} is equivalent to shifting the spectral parameter \bea z \rightarrow z - \alpha h^{\vee}f.\eea This effect is entirely analogous to the one found for Wilson lines in \cite{Costello:2017dso}.

\subsection{Anomaly Cancellation for Holomorphic Surface Defects}

\paragraph{} We can now formulate the anomaly cancellation result for holomorphic surface defects. 

\begin{figure}
\centering
\includegraphics[width=0.65\textwidth]{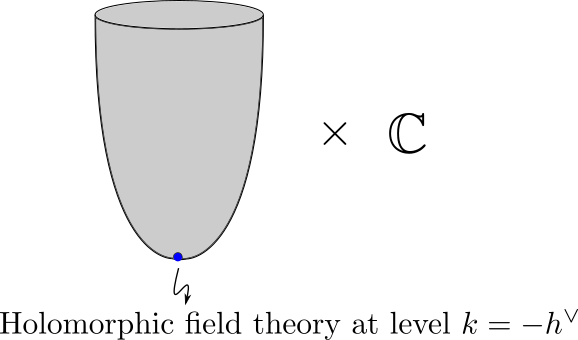}
\caption{The topological surface $\Sigma$ is the singular limit of a cigar geometry where the curvature is localized at the tip. In the bulk spacetime $\Sigma \times \mathbb{C}$ we have four-dimensional Chern-Simons theory. Since the curvature of the cigar $\Sigma$ is localized at the tip, this localizes the framing anomaly of 4d CS to $\{\text{tip} \} \times \mathbb{C}$. At the tip of the cigar, we couple to a holomorphic field theory at critical level. The framing anomaly of the bulk theory cancels the Kac-Moody anomaly of the holomorphic field theory.}  
\label{defect}
\end{figure}

\paragraph{} Suppose that we couple to a holomorphic field theory with Kac-Moody level $k$ at some point $w_0 \in \Sigma$. We have seen that the coupling of the bulk four-dimensional gauge field to the two-dimensional currents lead to the anomaly \bea \mathcal{A}_{\text{2d}} = \frac{k}{2\pi} \int \text{d}^2z \, \text{Tr} \big( \varepsilon\, \del_z A_{\bar{z}}|_{w_0} \big)\eea localized along $\{w_0\} \times \mathbb{C} \subset \Sigma \times \mathbb{C}$. On the other hand, the framing anomaly of the bulk four-dimensional Chern-Simons theory is \bea \mathcal{A}_{\text{4d}} = \frac{h^{\vee}}{8 \pi^2} \int_{\Sigma \times \mathbb{C}} \text{dvol}_{\Sigma} \, \text{d}^2 z \, R_{\Sigma} \text{Tr} \big( \varepsilon \del_z A_{\bar{z}} \big). \eea The four-dimensional framing anomaly begins to resemble the form of the two-dimensional Kac-Moody anomaly as the curvature of $\Sigma$ becomes more and more sharply localized at the insertion point $w_0$ of the defect. Suppose the Ricci curvature of $\Sigma$ is strictly localized at $w=w_0$ and takes the precise form \bea \sqrt{g} R_{\Sigma} = 4\pi \delta^{(2)}_{w=w_0}.\eea This is equivalent to saying that the Euler class $e(\Sigma)$ of $\Sigma$ is the two-form Poincare dual to the point $w_0 \in \Sigma$ \bea e(\Sigma) = \text{PD}[w_0]. \eea Then the two-dimensional and four-dimensional anomalies cancel \bea \mathcal{A}_{\text{2d}} + \mathcal{A}_{\text{4d}} = 0,\eea provided the Kac-Moody level is critical \bea k+h^{\vee} = 0. \eea Since \bea \chi(\Sigma) =  \int_{\Sigma} e(\Sigma) = 1,\eea $\Sigma$ is topologically a cigar. This can be made more explicit by letting $\Sigma = \mathbb{R}^2$ as a topological manifold, and taking $w_0$ to be the origin, and equipping it with a metric such that the curvature is a nascent delta function \bea \sqrt{g_{\epsilon}} R_{\epsilon} = 4 \pi \times \frac{1}{\pi \epsilon}e^{-\frac{x^2 + y^2}{\epsilon}}.\eea For finite $\epsilon$, this is a smooth cigar geometry, and in the $\epsilon \rightarrow 0$ limit it becomes singular at the origin with the Euler class becoming a delta function supported at the origin.

\paragraph{} In summary, we find that if the topological surface $\Sigma$ is the singular limit of a cigar geometry, we can insert a holomorphic field theory with $k=-h^{\vee}$ the tip. The result is a coupled 2d-4d system free of anomalies. This is summarized in Figure \ref{defect}. 

\section{Coadjoint Orbits and Monodromy Defects} \label{monodromy} Having shown that four-dimensional Chern-Simons theory can be coupled to holomorphic sigma models provided the Kac-Moody level is critical, we now discuss a particularly interesting defect that satisfies this criteria. This is the case of a holomorphic sigma model into a coadjoint orbit. Before discussing the general case, it is helpful to start with the example of $G= \text{SL}(2, \mathbb{C})$. 

\paragraph{} The orbit $X_j$ of the semisimple element $h_j := j h \in \mathfrak{sl}_2$ where $h = \begin{pmatrix}
1& 0 \\
0 & -1
\end{pmatrix}$ and $j$ is a non-zero complex number is the same as the variety in $\mathbb{C}^3$ given by the equation \bea Z_1^2 + Z_2^2 + Z_3^2 = j^2.\eea The Kostant-Kirilov two-form on coadjoint orbits makes $X_j$ into a holomorphic symplectic manifold. Explicitly, the symplectic form is given by \bea \Omega =  \frac{\text{d}Z_1 \wedge \text{d}Z_2}{\text{i} \,Z_3},\eea which is the natural complexification of the real symplectic two-form on the real two-sphere. The complex symplectic manifold $(X_j, \Omega)$ has $SL(2, \mathbb{C})$ as a subgroup of the space of symplectomorphisms. The vector fields that generate the $SL(2, \mathbb{C})$ action are the standard vector fields \bea K_i = \epsilon_{ijk}Z_j \frac{\del}{\del Z_k} \eea which satisfy \bea [K_i, K_j] = \epsilon_{ijk} K_k.\eea This is the Lie bracket relations on $\mathfrak{sl}_2$ written in the basis of anti-Hermitian matrices $e_i = \frac{1}{2 \text{i}} \sigma_i$ where $\sigma_i$ are the Pauli matrices. The moment maps in this basis are given by \bea \mu_i = \frac{1}{\text{i}} Z_i, \,\,\,\,\, i = 1,2, 3. \eea  The coadjoint orbit $X_j$ also admits an $I$-holomorphic vector field $U$ which scales the symplectic form \footnote{The space $X_j$ admits a hyperK\"{a}hler structure so that in addition to the two-form $\Omega$ we also have a non-degenerate, closed two-form $\omega_r$ which is of type $(1,1)$ in the complex structure $I$. With respect to this additional structure, $U$ is an $I$-holomorphic vector field which satisfies \bea \mathcal{L}_U \omega_r &=& 0, \\ \mathcal{L}_U \Omega &=& \text{i} \Omega.\eea} $$ \mathcal{L}_U \Omega = \text{i} \Omega$$ so that the holomorphic sigma model has $\mathbb{C}^*$ invariance. This will be more apparent after what we discuss below. 

\paragraph{} In order to show that the sigma model with target $X_j$ and symplectic form $\Omega$ can be coupled to four-dimensional Chern-Simons theory with $SL(2, \mathbb{C})$ gauge group, we must study the vertex algebra associated to this holomorphic sigma model, and in particular its Kac-Moody level. In order to do this, it is useful to find a description of $X_j$ as a cotangent bundle. It is clear that when literally stated like this, such a description is not possible since the complex symplectic form $\Omega$ on $X_j$ is not an exact two-form \footnote{One can see this by integrating $\Omega$ along the compact submanifold $$S^2 = \{(j \,x_1, j \,x_2, j \,x_3) \in X_j | (x_1,x_2, x_3) \in \mathbb{R}^3 \} \subset X_j.$$ In fact it is a generator of $H^2(X_j) = \mathbb{C}$. } and so it is not isomorphic to a cotangent bundle. However there is a description which is close enough. $X_j$ can be identified with an \textit{affine deformation} of $T^* \mathbb{P}^1$. Recall that $T^* \mathbb{P}^1$ can be covered with two patches $U$ and $V$ with local coordinates in each patch being written as $(\beta, \gamma)$ and $(\beta', \gamma')$. The standard gluing law across patches on $T^* \mathbb{P}^1$ is to say that \bea \gamma' = \frac{1}{\gamma}, \,\,\,\,\,\,\, \beta' = -\gamma^2 \beta.\eea The affine deformation of $T^*\mathbb{P}^1$ that we will identify with the coadjoint orbit $X_j$ consists of using a modified gluing law which identifies \bea \gamma'= \frac{1}{\gamma}, \,\,\,\,\,\,\ \beta' = -\gamma^2 \beta + 2j \gamma. \eea This is still a holomorphic change of coordinates across patches and thus still defines a valid complex two-manifold \footnote{Gluing laws which give affine deformations of the total space of the line bundle $\mathcal{O}(-2) \rightarrow \mathbb{P}^1$ are captured by the sheaf cohomology group $H^1\big( \mathbb{P}^1, \mathcal{O}(-2) \big)$. This is indeed one-dimensional, and for us is parametrized by the complex number $j$.}. The holomorphic symplectic form still takes the form \bea \Omega = \text{d}\beta \wedge \text{d} \gamma = \text{d}\beta' \wedge \text{d}\gamma'.\eea The crucial difference when $j\neq 0$ however is that the Liouville one-form $\Lambda = \beta \text{d}\gamma$, no longer has any global meaning, since in the other patch it takes the form \bea \Lambda' = \beta' \text{d}\gamma' - 2j\frac{\text{d}\gamma'}{\gamma'},\eea and is therefore singular at $\gamma' = 0$. Thus the affine deformation of $T^*\mathbb{P}^1$ by a non-zero $j$ is no longer an exact symplectic manifold. It is well-known that it is equivalent as a holomorphic symplectic manifold to the coadjoint orbit $X_j$. For completeness, we work out the argument that shows their equivalence. The action of $SL(2, \mathbb{C})$ on $ \mathbb{P}^1$ with the affine deformation by a non-zero $j$ is no longer a lift of the $SL(2, \mathbb{C})$ action on the base. Instead it is generated by the vector fields \bea \label{vece} K^e &=& \frac{\del}{\del \gamma}, \\ \label{vech} K^h &=& 2\beta \frac{\del }{\del \beta} - 2\gamma \frac{\del}{\del \gamma}. \\ \label{vecf} K^f &=& (-2j +2\beta \gamma) \frac{\del}{\del \beta} - \gamma^2 \frac{\del}{\del \gamma}. \eea One can check that these are globally well-defined vector fields on deformed $T^* \mathbb{P}^1$ and generate symplectomorphisms that moreover satisfy the $\mathfrak{sl}_2$ relations \bea [K_e, K_f] &=& K_h, \\ \,[K_h, K_e] &=& 2K_e, \\ \,[K_h, K_f] &=& -2K_f.\eea The corresponding moment maps written in the patch $U$ read \bea \label{mue}\mu^e &=& \beta, \\ \label{muh} \mu^h &=& 2j - 2 \beta \gamma, \\ \label{muf} \mu^f &=& 2j \gamma - \beta \gamma^2. \eea It can be checked that these remain holomorphic in the patch $V$. So we have holomorphic functions $\mu^e, \mu^f, \mu^h$ on the deformed cotangent bundle and moreover, these functions satisfy the quadratic identity \bea \frac{1}{4}(\mu^h)^2 + \mu^e \mu^f = j^2.\eea Therefore, letting \bea Z_1 &=& \frac{\mu^e + \mu^f}{2}, \\ Z_2 &=& \frac{\mu^e - \mu^f}{2 \text{i}}, \\ Z_3 &=& \frac{\mu^h}{2},\eea we find a holomorphic map $a: T^*\mathbb{P}^1 \rightarrow X_j$. Moreover, one can compute that the symplectic form $\Omega = \frac{\text{d}Z_1 \text{d}Z_2}{\text{i} \,Z_3}$ on $X_j$ indeed pulls back to the two-form $\text{d}\beta \wedge \text{d}\gamma$ on deformed $T^* \mathbb{P}^1$ \bea a^* \Big( \frac{\text{d}Z_1 \wedge \text{d}Z_2}{\text{i} \, Z_3} \Big) = \text{d} \beta \wedge \text{d} \gamma .\eea Thus the two spaces are equivalent as holomorphic symplectic manifolds. Moreover the isomorphism between the two manifolds is $\text{SL}(2, \mathbb{C})$-equivariant. This means that the  moment maps pull back in the right way.

\paragraph{} We have thus shown that there is an $\text{SL}(2, \mathbb{C})$-equivariant holomorphic symplectomorphism between $X_j$ and a particular deformation of $T^* \mathbb{P}^1$. The vertex algebra associated to $T^* \mathbb{P}^1$ on the other hand is known and was discussed previously in the paper. In order to study the vertex algebra associated to $X_j$, it is thus natural to look for deformations of the defining currents \eqref{p1e}-\eqref{p1f} which give rise to the vertex algebra associated to $T^* \mathbb{P}^1$. Such a deformation theory is well-understood and we recall the main features. We let $\chi(z)$ be an arbitrary Laurent series with complex coefficients. We take it to have spin $1$ so that we can write \bea \chi(z) = \sum_{n \in \mathbb{Z}} \frac{\chi_n}{z^{n+1}}. \eea The claim is that for every such $\chi$ we can deform the $\mathfrak{sl}_2$ currents as follows. First, given the local fields $\beta$ and $\gamma$, in the patch $U$ and $\beta', \gamma'$ in the patch $V$, we use the following gluing rule, the natural deformation of \eqref{gluegamma}, \eqref{gluebeta}: \bea \gamma' &=& \frac{1}{\gamma}, \\ \beta' &=& -( \beta(\gamma\gamma)) + 2 \del \gamma - (\chi \gamma) \eea for going between patches. Next we write the expression for the deformed local currents in the patch $U$. They read \bea e(z) &=& \beta(z), \\ h(z) &=& \chi(z) + 2(\gamma \beta)(z), \\ f(z) &=& -(\chi \gamma)(z) + 2 (\del \gamma)(z) - \big(\beta (\gamma \gamma) \big) (z).\eea For any $\chi(z)$ one can show that these currents satisfy the Kac-Moody algebra at level $k=-2$: \bea h(z) h(w) &\sim& \frac{-4}{(z-w)^2}, \\ h(z) e(w) &\sim& \frac{2e}{z-w}, \\ h(z) f(w) &\sim& -\frac{2f}{(z-w)}, \\ e(z)f(w) &\sim& \frac{-2}{(z-w)^2} + \frac{h(w)}{z-w}.\eea Therefore for any $\chi(z)$ we find a $\widehat{\mathfrak{sl}}_2$ algebra at the critical level. Moreover the rescaled stress tensor \bea S=(k+2)T = \frac{1}{4} (hh) + \frac{1}{2}(ef) + \frac{1}{2}(fe)\eea is such that the $\beta$ and $\gamma$ fields drop out, and can be expressed entirely in terms of $\chi$. It reads \bea \label{sintermsofchi} S = \frac{1}{4}\chi^2 - \frac{1}{2}\del \chi.\eea These are known as the Wakimoto current relations \cite{Wakimoto:1986gf}. The reader is refered to section 15.7 of \cite{DiFrancesco:1997nk} for a detailed proof of these relations. The vertex algebra associated to $X_j$ comes about when we specialize the value of $\chi$ to be \bea \chi_j = 2j.\eea In particular, this means that the rescaled Sugawara current becomes \bea S(z) = j^2.\eea This is a vertex algebra manifestation of the equation $$\frac{1}{4} \mu_h^2 + \mu_e \mu_f = j^2.$$ The vertex algebra associated to $X_j$ is thus isomorphic to the vacuum module $V_{-2}(\widehat{ \mathfrak{sl}}_2)$ modulo the center with central character $\chi = 2j$. In particular, this involves setting the singular vector \bea \Big(\frac{1}{4} h_{-1}h_{-1} + \frac{1}{2} e_{-1}f_{-1} + \frac{1}{2} f_{-1}e_{-1}-j^2 \Big)|0 \rangle = 0 .\eea whereas all other singular vectors are set to vanish.  In particular, the vertex algebra associated to $X_j$ is still at the critical level, and we can therefore conclude that the system can be coupled consistently to four-dimensional Chern-Simons theory.

\paragraph{} Although we have discussed the case of $\mathfrak{sl}_2$ in detail, there is a version of these results that hold for an arbitrary $\mathfrak{g}$. Suppose we consider the coadjoint orbit of a regular, semi-simple element $\alpha$ in a complex Lie group $G$. The semi-simple element $\alpha$ can be taken to be in the (dual) of the Cartan $\mathfrak{h}^{\vee}$ without loss of generality. It is then well-known that there is an affine deformation of the symplectic manifold $T^*(G/B)$ where $B$ denotes the Borel subgroup of $G$ which is symplectomorphic to the coadjoint orbit $\text{Orb}(\alpha)$. The deformation can be viewed as follows. It is well-known that the Dolbeault cohomology group of $(1,1)$ forms on the K\"{a}hler manifold $G/B$ is \bea H^{(1,1)} \big( G/B \big) \cong \mathfrak{h}.\eea By the Cech-Dolbeault isomorphism we then have the sheaf cohomology group \bea H^{1}\big(G/B, T^*(G/B) \big) \cong \mathfrak{h},\eea which corresponds precisely to affine deformations of the cotangent bundle $T^* (G/B)$. One can also give more explicit formulas in terms of gluing rules, but we will not do so. It is also known in a similar way that the $\widehat{\mathfrak{g}}$-currents associated to $T^* (G/B)$ can be deformed by any element \bea \chi \in \mathfrak{h} \otimes \mathbb{C}((z)) \eea while remaining at the critical level $k = -h^{\vee}$. The vertex algebra associated to $\text{Orb}(\alpha)$ arises upon specialization to \bea \chi = \alpha .\eea In particular, it remains critical.

\paragraph{} Having demonstrated that four-dimensional Chern-Simons can be consistently coupled to the holomorphic sigma model onto the coadjoint orbit $X_{\alpha}$, we now show that this defect is equivalent to what is commonly known as a ``monodromy defect". Monodromy defects in a given gauge theory with gauge group $G$ are codimension two defects such that the monodromy of the gauge field along any path that goes around the defect is in a fixed conjugacy class of $G$. Codimension two sigma models with target spaces being coadjoint orbits are well-known to give microscopic descriptions of such monodromy defects in particular examples. Two known cases where this holds is in three-dimensional Chern-Simons theory, and four dimensional $\mathcal{N}=4$ supersymmetric gauge theory. For the standard three-dimensional Chern-Simons theory, the codimension two defect is simply a one-dimensonal sigma model consisting of gauged topological quantum mechanics with target space being a coadjoint orbit of the real gauge group. Via the Borel-Weil-Bott theorem, this is just an alternate description of a Wilson line. In $\mathcal{N}=4$ Yang-Mills theory, on the other hand, monodromy defects were studied in \cite{Gukov:2006jk}, and it was shown that their microscopic description is closely related to the two-dimensional $\mathcal{N}=(4,4)$ hyperK\"{a}hler sigma model into a complex coadjoint orbit \cite{Gukov:2008sn}. We now show that four-dimensional Chern-Simons theory coupled to the holomorphic sigma model into the coadjoint orbit is equivalent to a monodromy defect.

\paragraph{} Recall that the action of four-dimensional Chern-Simons theory with gauge algebra $\mathfrak{g}$ coupled to a holomorphic sigma model into $(X, \Omega)$ reads (once again we let $x,y$ be standard real coordinates in the topological direction, and $z, \bar{z}$ coordinates in the holomorphic direction) \bea \label{2d4daction} S= \frac{1}{\pi} \int \text{d}^2 z \text{d}^2 x \, \kappa_{ab} \big(A_y^a \del_{\bar{z}} A^b_x + A^a_{\bar{z}} F^b_{xy} \big)- \frac{1}{2\pi}\int \text{d}^2 z \big( \Lambda_i \del_{\bar{z}} \phi^i + \mu_a \,A^a_{\bar{z}}|_{(x,y)=(0,0)}  \big). \eea The equations of motion for the combined system read as follows. The gauge field satisfies \bea \label{xycurvature} F^a_{xy} &=& \frac{1}{2}\kappa^{ab} \mu_b \, \delta^{(2)}(x, y). \eea along with \bea F^a_{x \bar{z}} = F^a_{y \bar{z}} =0, \eea  whereas the sigma model field satisfies \bea D_{\bar{z}} \phi^i = 0 \eea where \bea D_{\bar{z}} \phi^i = \del_{\bar{z}} \phi^i - A^a_{\bar{z}} K_a^i. \eea The monodromy around the origin $(x,y) = (0,0)$ in the topological plane is then immediately computed to be \bea \mathbb{M} = \text{exp} \Big( \frac{1}{2} \kappa^{ab} \mu_a t_b \Big) \in G. \eea For a generic $X$ this has no reason to be in a fixed conjugacy class as the sigma model field $\phi$, and thus the moment map $\mu$ varies.  We now specialize to the situation which is the exception. Let $\mathfrak{g}^{\vee}$ be the linear dual of the Lie algebra $\mathfrak{g}$ and suppose we take $X$ to be the coadjoint orbit $X_{\alpha}$ for a regular, semi-simple element $\alpha \in \mathfrak{h}^{\vee}$ in the dual of the Cartan. $X_{\alpha}$ is naturally a subset of $\mathfrak{g}^{\vee}$. With the natural Kostant-Kirilov holomorphic symplectic form on $X_{\alpha}$, the moment map \bea \mu: X_{\alpha} \rightarrow \mathfrak{g}^{\vee} \eea is simply the embedding map. By definition every point in $X_{\alpha}$ is obtained by conjugating the fixed element $\alpha$ by some element of $G$. This implies that the conjugacy class of $\mathbb{M} = \text{exp} \Big( \frac{1}{2} \kappa^{ab} \mu_a t_b \Big)$ is simply the conjugacy class (writing $\alpha = \alpha_a t^a$) \bea \mathbf{C}_{\alpha} = \{ \text{exp} \Big( \frac{1}{2} \kappa^{ab} \alpha_a t_b \Big) \}. \eea It is instructive to demonstrate the claim that the monodromy of the gauge field for a path around $(x,y) = (0,0)$ is in a fixed conjugacy class more explicitly for the gauge group $SL(2, \mathbb{C})$. We will do this using both descriptions of the coadjoint orbit. First for $j \neq 0$, we use the direct description as the complexified sphere. Recall that the moment maps in the basis $e_i = \frac{1}{2 \text{i}} \sigma_i$ of $\mathfrak{sl}_2$ were given by $\mu_i = \frac{1}{i} Z_i$, for $i=1,2,3$. The Killing form in this basis is simply \bea \kappa = -\frac{1}{2} \, \text{id},\eea where $\text{id}$ denotes the rank $3$ identity matrix. This means that the $xy$-component of the curvature is the $\mathfrak{sl}_2$ element \bea \kappa^{ab} \mu_a e_b\eea times a $\delta$-function. Therefore we have \bea F_{xy} = \frac{1}{2} \delta^{(2)}(x,y)\begin{pmatrix}
Z_3 & Z_1 - \text{i} Z_2 \\
Z_1 + \text{i} Z_3 & -Z_3
\end{pmatrix} .\eea Therefore the monodromy $\mathbb{M}$ along a path going around $(x,y)=(0,0)$ is given by \bea \mathbb{M} = \text{exp} \, \frac{1}{2} \begin{pmatrix}
Z_3 & Z_1 - \text{i} Z_2 \\
Z_1 + \text{i} Z_3 & -Z_3
\end{pmatrix}.  \eea Because of the relation $$Z_1^2 + Z_2^3 + Z_3^2 = j^2,$$ this is conjugate to the element \bea \mathbf{C}_j = \begin{pmatrix}
\text{exp}(\frac{j}{2}) & 0 \\ 
0 & \text{exp}(-\frac{j}{2})
\end{pmatrix}. \eea Therefore we conclude that the defect for $j \neq 0$ implies that the monodromy of the gauge field around the defect is in the semi-simple conjugacy class of the element $\text{diag}(\alpha, \alpha^{-1})$ where \bea \alpha = e^{\frac{j}{2}}.\eea We can also arrive at the same conclusion using the description of the defect as a sigma model into the deformed cotangent bundle $T^* \mathbb{P}^1$. Here, we use the moment maps given in \eqref{mue}-\eqref{muf} along with the standard $\mathfrak{sl}_2$ basis $\{e, h, f \}$ in which the non-zero Killing form elements read \bea \kappa(e,f) = 1, \,\,\,\, \kappa(h,h) = 2.\eea The curvature $F_{xy}$ is then equal to the element \bea \frac{1}{2} \big( \kappa(e,f)^{-1} (\mu_e f + \mu_f e) +  \kappa(h,h)^{-1} \mu_h h \big).\eea Writing it out explicitly gives \bea F_{xy} = \frac{1}{2} \delta^{(2)}(x,y) \begin{pmatrix} j-\beta \gamma & 2j\gamma - \beta\gamma^2 \\ \beta & -j + \beta \gamma \end{pmatrix}. \eea Once again because of the relation $$\frac{1}{4} \mu_h^2 + \mu_e \mu_f = j^2,$$ any connection whose curvature satisfies such an equation has a monodromy in the conjugacy class $\mathbf{C}_j$ as long as $j \neq 0$. The advantage of this current description is that it continues to make sense at $j=0$. Here we are simply coupling to the ordinary cotangent bundle $T^* \mathbb{P}^1$. At $j=0$ the curvature is now a $\delta$-function times the degenerate matrix \bea \frac{1}{2} \begin{pmatrix}
-\beta \gamma & -\beta \gamma^2 \\ 
\beta & \beta \gamma
\end{pmatrix} .\eea We therefore conclude that at $j=0$ the monodromy lies in the unipotent conjugacy class of elements conjugate to \bea \mathbf{C}_{j=0} = \begin{pmatrix}
1 & 1 \\ 
0 & 1 
\end{pmatrix}. \eea Geometrically this makes sense, since it is indeed well-known that $T^* \mathbb{P}^1$ provides a resolution of the singular space $X_e$, the coadjoint orbit of the nilpotent element $e$ in $SL(2, \mathbb{C})$. 

\subsection{Description as a Disorder Defect} So far we have discussed one version of the monodromy defect where we couple the four-dimensional Chern-Simons theory to a sigma model into a deformed cotangent bundle of a real coadjoint orbit (equivalently a sigma model into a complex coadjoint orbit). We showed that for such a coupled system, the monodromy of the gauge field around the defect lies in a fixed conjugacy class dictated by the choice of orbit. We now give a description that uses no sigma model at all. Instead it will be a codimension two disorder operator supported along the holomorphic plane. 

\paragraph{} Typically disorder defects supported along a submanifold $Y$ of the ambient spacetime $X$ arise by first studying a model solution of the equations of motion on $X \backslash Y$ which has a certain type of singularity as we approach the defect locus $Y$. The defect is then defined by saying that we study the quantum theory on a space of field configurations on $X \backslash Y$ such that as we approach $Y$, the fields in our field space approach the singular model solution. 

\paragraph{} For four-dimensional Chern-Simons theory, we can easily work out a model solution of the equations of motion \bea F_{xy} = F_{x\bar{z}} = F_{y \bar{z}} = 0\eea which is singular along the locus $$ Y = \{(x,y)=(0,0)\} \subset \mathbb{R}^2 \times \mathbb{C}.$$  Let \bea \zeta : \mathbb{C} \rightarrow \mathfrak{h} \subset \mathfrak{g} \eea be a \underline{regular}\footnote{Recall that an element $\alpha \in \mathfrak{h}$ is called regular if its centralizer is $\mathfrak{h}$. A holomorphic function $\zeta : \mathbb{C} \rightarrow \mathfrak{h}$ will be called regular if $\zeta(z)$ is regular for all $z$. } holomorphic function valued in the Cartan subalgebra of $\mathfrak{g}$. Let $\theta$ denote the standard angular coordinate on the topological $(x,y)$-plane. Then \bea A =  \zeta(z) \text{d} \theta + \mathfrak{t}(z,\bar{z}) \text{d}\bar{z}\eea is a solution of the equations of motion for any $\mathfrak{h}$-valued connection $$\mathfrak{t}(z, \bar{z}) \text{d}\bar{z}$$ on $\mathbb{C}$. Moreover, this has a singularity as we approach the origin in the topological plane, because the one-form $\text{d}\theta$ is singular at the origin. It is also rotationally invariant along the topological plane. 

\paragraph{} We can then define a defect associated to a regular $\mathfrak{h}$-valued holomorphic function $\zeta(z)$ by specifying the following space of field configurations. We consider partial $G$-connections \bea A_x \text{d}x + A_y \text{d}y + A_{\bar{z}} \text{d}\bar{z} \eea on the space $\mathbb{R}^2 \times \mathbb{C} \backslash \{(x,y)= (0,0) \}$ such that \bea A_x \text{d}x + A_y \text{d}y \rightarrow \zeta(z) \text{d} \theta \,\,\, \text{ as } \,\,\, (x,y) \rightarrow (0,0)\eea and the limit of $A_{\bar{z}} \text{d}\bar{z}$ is well-defined as $(x,y) \rightarrow (0,0)$ in which it becomes an $\mathfrak{h}$-valued connection on $\mathbb{C}$ \bea \text{lim}_{(x,y) \rightarrow (0,0) } \, A_{\bar{z}} \text{d}\bar{z} \, \in  \,\{ T\text{-connections on } \mathbb{C}  \},\eea where $\mathfrak{h} = \text{Lie}(T)$. The space of fields is acted on by the group of gauge transformations $\mathcal{G}$ consisting of maps $$g: \mathbb{R}^2 \times \mathbb{C} \backslash \{(x,y)= (0,0) \} \rightarrow G$$ such that the limit as $(x,y) \rightarrow (0,0)$ is well-defined and is such that $g$ becomes valued in the Cartan torus $T$ \bea \text{lim}_{x,y \rightarrow (0,0)} \,  g \, \in \, \text{Map}(\mathbb{C}, T). \eea We note that the data along the defect locus is purely holomorphic: it consists of an $\mathfrak{h}$-valued holomorphic function, and a space of field configurations being $\mathfrak{h}$-valued holomorphic bundles on $\mathbb{C}$. Thus our disorder defect is holomorphic.

\paragraph{} We now show that the description of the codimension two disorder defect, and the sigma model description (generalized in an appropriate, important way) are equivalent.  We explain the equivalence in detail for our favorite gauge group $G= \text{SL}(2, \mathbb{C})$. Let us begin with the description of the defect as a sigma model into the coadjoint orbit $X_j$. Here we will need the generalization alluded to briefly in Section 2: the sigma model into $X_j$ makes sense even when $j$ is promoted from a fixed complex number to a promoted to a nowhere vanishing holomorphic function on $\mathbb{C}$ \bea j \rightarrow j(z) .\eea More precisely, what we are doing is the following. By rescaling the coordinates $Z_i$, we can consider a fixed complex manifold $X$ defined by the equation $Y_1^2 + Y_2^2 + Y_3^2 = 1$. We consider the holomorphically varying holomorphic symplectic form \bea \Omega(z) = j(z)\frac{\text{d}Y_1 \wedge \text{d}Y_2}{\text{i} Y_3}.\eea In terms of the language introduced in Section 2, this corresponds to choosing the $(3,0)$ form \bea \mathbb{T } = \frac{j(z) \text{d}z \wedge \text{d}Y_1 \wedge \text{d}Y_3 }{\text{i} Y_3} \eea on $\mathbb{C} \times X$. This equation suggests that is natural to think of $j(z) \text{d}z$ as a one-form on $\mathbb{C}$. Letting $g(z)$ be such that $g'(z) = j(z)$ we can write the action of the theory explicitly as \bea S = -\frac{1}{2\pi \text{i}} \int_{\mathbb{C}}g(z)\phi^* \Big( \frac{\text{d}Y_1 \wedge \text{d}Y_2}{Y_3} \Big). \eea Our discussion of the deformation of the current algebra by the field $\chi(z)$ shows that the level is critical for all $j(z)$ and so we can still couple our holomorphic field theory to four-dimensional Chern-Simons theory (the precise relation between $\chi$ and $j$ is that $\chi(z) = 2j(z)$).

\paragraph{} We now study the equations of motion of our coupled 2d-4d system: they are \bea F_{xy} &=& \frac{1}{2} \begin{pmatrix}
Z_3 & Z_1 - \text{i} Z_2 \\ 
Z_1 + \text{i}Z_2 & Z_3
\end{pmatrix} \delta^{(2)}(x,y) , \\ F_{x\bar{z}} &=& F_{y\bar{z}} = 0, \\ 0&=& \del_{\bar{z}} Z_i + \epsilon_{ijk} A^j_{\bar{z}} Z_k , \,\,\,\, i=1,2,3, \eea where the sigma model fields $Z_1, Z_2, Z_3$ are subject to the constraint \bea Z_1^2 +Z_2^2 + Z_3^2 = j^2(z).\eea We can try to solve these equations in a convenient gauge. By using an $\text{SL}(2,\mathbb{C})$ gauge transformation varying only along $\mathbb{C}$ we can go to a gauge where \bea Z_1 = Z_2 = 0, \,\,\, Z_3 = j(z).\eea The sigma model equation of motion for $i=3$ automatically holds, whereas the $i=1,2$ equations imply that the gauge field $A_{\bar{z}}$ must satisfy \bea A_{\bar{z}}^1|_{x=y=0} = A_{\bar{z}}^2|_{x=y=0} = 0.\eea The remaining equation of motion for the gauge field along the $(x,y)$-directions is then \bea \label{eomfxy}  F_{xy} = \frac{1}{2} j(z) \sigma^{3} \delta^{(2)}(x,y), \eea which subjects any solution (in a rotationally invariant gauge along $\mathbb{R}^2$) to the singular behavior \bea A_x \text{d}x + A_y \text{d}y = \frac{1}{4 \pi} j(z) \text{d} \theta + \dots.\eea In summary we have found that the sigma model fields can be gauged away provided along the defect locus $A_{\bar{z}}$ becomes a Cartan valued connection, and the connection $A_x \text{d}x + A_y \text{d}y$ has singular behavior precisely of the required sort with the identification \bea \zeta(z) = \frac{1}{4\pi} j(z) \sigma^{3}. \eea This was precisely the field space of the disorder operator. Moreover, because we had to use a gauge transformation varying along $\mathbb{C}$ in order to gauge fix the $X$-valued fields, the remaining gauge group is such that along the defect locus, the gauge transformations can only be diagonal, namely they are valued in the Cartan of $\text{SL}(2, \mathbb{C})$. 

\paragraph{} It is also instructive to reproduce the same field space from the description of the sigma model as deformed $T^* \mathbb{P}^1$. Once again, here we are talking about a deformation with a spatially varying parameter $j(z)$. The equations of motion for the gauge field in this description are \bea F_{xy} &=& \frac{1}{2} \begin{pmatrix}
j(z) - \beta \gamma & 2j(z) \beta  - \beta \gamma^2 \\ 
\beta & -j(z)  + \beta \gamma
\end{pmatrix} , \\ F_{x\bar{z}} &=& F_{y\bar{z}} = 0,\eea whereas the sigma model equations (using the explicit vector fields \eqref{vece}-\eqref{vech}) are  \bea \del_{\bar{z}} \gamma - A^e_{\bar{z}} + A^f_{\bar{z}}\gamma^2 + 2A^h_{\bar{z}} \gamma &=& 0, \\ \del_{\bar{z}} \beta - 2A^h_{\bar{z}} \beta + 2A^f_{\bar{z}} (j(z) - \beta \gamma) &=& 0. \eea In order to solve the equations, we again make a convenient gauge choice. The gauge transformations act infinitesimally via \bea \delta \gamma &=& \varepsilon^{e} + \varepsilon^{h}(-2\gamma) + \varepsilon^{f}(-\gamma^2) , \\ \delta \beta &=& \varepsilon^{h}(2\beta) + \varepsilon^{f}(-2j + 2\beta \gamma). \eea We can use the gauge transformation parameter $\varepsilon^{e}$ to go to a gauge with $\gamma = 0$. The equation of motion involving $\gamma$ in this gauge is then requires \bea A^a_{\bar{z}}|_{(x,y)=(0,0)} = 0. \eea On the other hand, in the gauge $\gamma=0$, the infinitesimal gauge transformation in the $f$-direction acts via \bea \delta_f \beta =  -2j(z) \varepsilon^f. \eea From this it becomes clear that by using the $f$-gauge transformation, we can also go to a gauge where $\beta = 0$ (here the assumption that $j(z)$ is nowhere vanishing becomes crucial). The equation of motion involving $\beta$ then requires \bea A^f_{\bar{z}} |_{(x,y) = (0,0)} = 0.\eea Thus we have found the condition that along the defect locus, $A_{\bar{z}}$ becomes Cartan-valued. Moreover, the equations for the gauge field in the gauge $\gamma = \beta = 0$ reduce to the ones \eqref{eomfxy}. We reproduce the same result from either description. There is a similar analysis that can be done to show the equivalence for any gauge algebra $\mathfrak{g}$.

\paragraph{} Generalizing the analysis to a situation when $j(z)$ can have zeros is an interesting issue that will require us to incorporate conjugacy classes of nilpotent elements of $\mathfrak{g}$. We postpone this to future work.  

\section{Conclusions} In this paper we have discussed an anomaly cancellation mechanism that allows us to couple four-dimensional Chern-Simons theory to holomorphic field theories with global symmetry. In particular, coupling to holomorphic field theories at the critical level is allowed, provided the topological surface has a curvature sharply localized at the insertion point of the defect. 

\paragraph{} We discussed the equivalence between holomorphic sigma models into complex coadjoint orbits, and a certain type of disorder operator which constrains the singular behavior of the gauge field. Here we encountered a novel generalization of the standard sort of monodromy defect discussed in the literature: in our setup the conjugacy class of the monodromy around the defect is not fixed, but is allowed to vary holomorphically according to a holomorphic function that is specified in advance. It is therefore appropriate to dub the class of defects we discussed in this paper as \textit{holomorphic monodromy defects}. 

\paragraph{} A given holomorphic monodromy defect is specified by a holomorphic function $\zeta: \mathbb{C} \rightarrow \mathfrak{h}$ and in this paper we have considered only the simplest case where the function is regular everywhere. Incorporating zeros (more generally, non-regular loci) and poles is an interesting issue we hope to address in the future. Whereas at zeroes, one will have to incorporate unipotent conjugacy classes, poles are more subtle because they can lead to a failure of gauge invariance. A solution we wish to expand upon in \cite{CIKY} is that poles are naturally associated to endpoints of Wilson lines that terminate on the defect. Ultimately, one would like to give the Bethe Ansatz equations and the Bethe eigenstates of integrable models a natural and direct interpretation in four-dimensional Chern-Simons theory. We believe that monodromy defects will play an important role in this endeavor.

\end{document}